\documentclass[useAMS,usenatbib]{mnras}

\usepackage{
amsmath,
amssymb,
booktabs,
graphicx,
xspace,
hyperref,
xcolor,
}
\hypersetup{draft}

\newcommand\eg{e.\,g.\xspace}
\newcommand\ie{i.\,e.\xspace}
\newcommand\cf{c.\,f.\xspace}
\def\spose#1{\hbox to 0pt{#1\hss}}
\newcommand\lta{\mathrel{\spose{\lower 3pt\hbox{$\mathchar"218$}}
     \raise 2.0pt\hbox{$\mathchar"13C$}}}
\newcommand\gta{\mathrel{\spose{\lower 3pt\hbox{$\mathchar"218$}}
     \raise 2.0pt\hbox{$\mathchar"13E$}}}
\newcommand\kpc{\ensuremath{\mathrm{kpc}}}
\newcommand\pc{\ensuremath{\mathrm{pc}}}
\newcommand\gyr{\ensuremath{\mathrm{Gyr}}}
\newcommand\msun{\ensuremath{\mathrm{M}_\odot}\xspace}
\newcommand\rb{\ensuremath{R_\mathrm{b}}}
\newcommand\jzjc{\ensuremath{\frac{j_z}{j_\mathrm{c}\left(e\right)}}\xspace}
\newcommand\om[1]{\ensuremath{\Omega_\mathrm{#1}}}

\def\changed#1{#1}

\newcommand\psb{\citepalias{Sellwood2002}\xspace}
\newcommand\tsb{\citetalias{Sellwood2002}\xspace}
\newcommand\thfifteen{\citetalias{Herpich2015}\xspace}
\newcommand\phfifteen{\citepalias{Herpich2015}\xspace}
\newcommand\olr{OLR\xspace}
\newcommand\corot{CR\xspace}

\def\apj{ApJ}
\def\aj{AJ}
\def\mnras{MNRAS}
\def\aap{A\&A}

\def\apjl{ApJ}
\def\aaps{A\&AS}

\newcommand\sortbib[1]{}
\defcitealias{Herpich2015}{paper I}
\defcitealias{Sellwood2002}{SB02}

\title[Formation of anti-truncations]{How to bend galaxy disc profiles II: stars surfing the
bar in Type-III discs}
\author[Herpich et al.]{
    J. Herpich$^{1}$%
\thanks{E-Mail: jakob@jkherpich.de}%
\thanks{Member of the International Max Planck Research School for
    Astronomy and Cosmic Physics at the University of Heidelberg,
    IMPRS-HD, Germany.
    },
    G.\,S. Stinson$^{1}$,
    H.-W. Rix$^{1}$,
    M. Martig$^{1}$,
    A.\,A. Dutton$^{1,2}$%
\\
$^{1}$Max-Planck-Institut für Astronomie, Königstuhl 17, 69117, Heidelberg, Germany\\
$^{2}$New York University Abu Dhabi, PO Box 129188, Abu Dhabi, UAE}
\begin{document}

\maketitle

\begin{abstract}
The radial profiles of stars in disc galaxies are observed to be either purely
exponential (Type-I), truncated (Type-II) or anti-truncated (Type-III) exponentials.
Controlled formation simulations of isolated galaxies can reproduce all of these
profile types by varying a single parameter, the initial halo spin.
In this paper we examine these simulations in more detail in an effort to identify
the physical mechanism that leads to the formation of Type-III profiles.
The stars in the anti-truncated outskirts of such discs are now on eccentric
orbits, but were born on near-circular orbits at much smaller radii.
We show that, and explain how, they were driven to the outskirts via non-linear
interactions with a strong and long-lived central bar, which greatly boosted their
semi-major axis but also their eccentricity.
While bars have been known to cause radial heating and outward migration to
stellar orbits, we link this effect to the formation of Type-III profiles.
This predicts that the anti-truncated parts of galaxies have unusual kinematics for
disc-like stellar configurations: high radial velocity dispersions and slow net rotation.
Whether such discs exist in nature, can be tested by future observations.

\end{abstract}
\begin{keywords}
galaxies: kinematics and dynamics -- galaxies: formation -- galaxies: structure -- methods: numerical
\end{keywords}

\section{Introduction}
\label{sec:intro}
\subsection{Observed disc breaks}
\label{sec:observed_breaks}
Most late-type galaxies share one common property: an approximately exponential
stellar surface density
profile \citep{deVaucouleurs1957, deVaucouleurs1958, Freeman1970} which extends over several
scale lengths.
As observations reached fainter surface brightness limits,
a more diverse picture of late-type morphologies has become apparent:
a major portion of such galaxies actually exhibit broken exponential profiles
\citep[\eg][]{vanderKruit1979, Pohlen2004}.
First, truncated or down-bending disc profiles were discovered
\citep[\eg][]{Pohlen2002}.
These discs' profiles feature a break at a radius outside of which the surface brightness profile
is still exponential, but steeper.
By contrast, \citet{Erwin2005} observed anti-truncated or up-bending profiles in barred
S0-Sb galaxies.
The breaks in these galaxies have the opposite behaviour: the slope of the exponential profile
outside of the break is shallower than its inner counterpart.
Thus stellar disc profiles fall into three different categories:
\begin{description}
\item[Type-I:] pure exponential profiles
\item[Type-II:] truncated or down-bending profiles
\item[Type-III:] anti-truncated or up-bending profiles.
\end{description}

Several studies investigated the abundance of each type of disc breaks in different
observational samples \citep[\eg][]{Pohlen2006, Erwin2008, Gutierrez2011, Maltby2012a,
Maltby2015} with the result that the relative abundances vary with morphological type.
\citet{Pohlen2006} found that pure exponential (Type-I) discs are rather rare (10 \%).
\citet{Head2015} find that more than 70\,\% of the galaxies with broken disc profiles
in the
Coma Cluster have a bar, despite a low bar fraction (about 20 \%) in the overall
sample.
They therefore relate the origin of Type-III profiles to bars.
A similar statement has been made by \citet{Kim2014} for Type-II profiles in
galaxies with small bulges.
Another possible cause for Type-III profiles, dubbed \emph{disc fading}, are stellar
halos which outshine the stellar disc at large radii,
although \citet{Maltby2012a, Maltby2015} found that this can only explain a minor fraction of
observed Type-III light profiles.

\subsection{Theoretical explanations of breaks}
\label{sec:explained_breaks}
The formation of stellar Type-II disc profiles has been studied in numerical
models of isolated galaxies \citep[\eg][]{Debattista2006, Roskar2008, Foyle2008,
Minchev2012}.
Simulations \citep{Roskar2008} and analytical models \citep{Dutton2009} suggest that
the size of the inner disc, \ie the radial position of the disc break, is
related to an upper limit of gas angular momentum in a rotationally supported star
forming gaseous disc.
In this picture, stars that populate the regions outside the break radius must have been
born inside the break.
\citet{Roskar2008} showed that radial migration, induced by transient spiral arms
\citep[first described by][hereafter \tsb]{Sellwood2002},
can move a sufficient number of stars outside the break radius and populate that
region.
\changed{Stars that migrate via this mode are heavily biased to be on near-circular rather than
eccentric orbits and to be vertically cool
\citep[\tsb;][]{Roskar2012,Solway2012,Vera-Ciro2014}.
An important diagnostic of this mechanism is that the migrated stars preserve their circularity
and, hence, the majority of the migrating stars stay on near-circular orbits
\citep[see also \tsb;][]{Roskar2012,Solway2012,Vera-Ciro2014}.}
This mechanism predicts positive stellar age gradients outside the break in Type-II
discs \citep{Roskar2008}.
The existence of such age profiles has been confirmed by observed colour
profiles \citep{Azzollini2008,Bakos2008}, spectroscopic observations of NGC 6155
\citep{Yoachim2010} and by resolved stellar populations \citep{Radburn-Smith2012}.
Recent work by \citet{RuizLara2015} found that positive age gradients in galaxy outskirts
are not unique to Type-II discs.
They claim that these age profiles are not linked to Type-II profiles but do
not rule out stellar migration as a possible mechanism to populate the regions
beyond the disc break.

The origin of Type-III disc profiles is still poorly understood.
Most of the proposed mechanisms resort to external forces \citep[\eg][]{Younger2007,
Kazantzidis2009, Roediger2012, Borlaff2014}.
However, \citet{Maltby2012} and \citet{Head2015} found no dependence of a galaxy's disc
break type on its environment.
This disfavours an external origin of breaks in stellar discs.
The first successful attempt to produce Type-III disc profiles in simulations
of isolated galaxies was put forward in a preceding letter
\citep[\thfifteen hereafter]{Herpich2015}.

\subsection{Halo spin disc break correlation}
\label{sec:spin_breaks}

In \thfifteen, we analyse a set of controlled simulations of galaxy formation
in Milky-Way mass halos with varying initial angular momentum.
We found that the type of the profile break
correlates with initial host halo spin.
Low-spin halos produce Type-III disc profiles, high-spin halos yield
Type-II profiles.
In a sharp transition between these two regimes (intermediate spin) pure exponential
Type-I
profiles formed.

In the present paper we present results of a more thorough analysis of the low-spin simulations
that were the subject of \thfifteen, \ie the simulations that formed
Type-III profiles.
Thus we will only briefly review the properties of the simulations and refer the reader to
\thfifteen for a more detailed description.

In this paper we seek a more analytic understanding of how Type-III profiles may form.
We start out with a short overview of previous theoretical work on stellar dynamics in
non-axisymmetric rotating potential perturbations such as spiral patterns or bars
(section \ref{sec:dynamics}).
In section \ref{sec:simulations} we quickly review the simulations from \thfifteen.
The simulations with the lowest spin will be analysed in more detail and with a
particular focus on stellar orbit evolution and bar properties in section \ref{sec:results}.
There we also identify the physical cause for the formation of Type-III profiles
and illustrate it with a simple toy model in section \ref{sec:toy_model}.
Observational signatures of such Type-III profiles are presented in section
\ref{sec:observations}.
Finally, we summarize and discuss our results in section \ref{sec:discussion}.

\section{Dynamics in stellar discs}
\label{sec:dynamics}

In their landmark work, \tsb show that non-axisymmetric perturbations in the
potential (\eg spiral arms or bars) can drive radial migration of stars in stellar discs.
\citet{Binney2008} describe the respective physics in sections 3.3.2 and 3.3.3
which we briefly summarize here.

In such a potential Jacobi's integral $e_\mathrm{J}$ is an integral
of motion, \ie it is constant.
It is defined as
\begin{equation}
e_\mathrm{J}=e-\om pj_z,
\label{eq:jacobi_energy}
\end{equation}
where $e$ is a test particle's specific orbital energy and $j_z$ is the $z$-component of the
specific angular momentum in a non-rotating frame%
\footnote{We implicitly assume the perturbation to rotate about the
$z$-axis.}.
If $e_\mathrm{J}$ and \om p are constant, equation (\ref{eq:jacobi_energy}) implies that
changes in energy $e$ are proportional to changes in angular momentum $j_z$:
\begin{equation}
\Delta e = \om p\Delta j_z
\label{eq:delta_e}.
\end{equation}
It is conceptually useful to plot $e$ versus $j_z$.
That plot is commonly referred to as a \emph{Lindblad diagram}.
Given these idealized conditions, equation \eqref{eq:delta_e} governs the trajectory of
test particles in the Lindblad diagram: it must have a slope that equals the pattern
speed \om p.
(See Fig. 1 in \tsb.)

Trajectories along other slopes in the Lindblad diagram are
possible, but such motions cannot be caused by perturbations rotating with a constant
pattern speed, such as bars.

\tsb also find that to first order changes in the radial action $J_R$ are
proportional to the difference between the pattern speed $\om p$ and the azimuthal angular
velocity of the star's guiding centre $\om\star$:
\begin{equation}
    \Delta J_R/m_\star=\frac{\om p - \om\star}{\omega_R}\Delta j_z
    \label{eq:heating_radial_migration}
\end{equation}
where $\omega_R$ and $m_\star$ are the radial frequency and mass of the star respectively.
Consequently, changes in angular momentum at the corotation resonance
($\om p = \om\star$, hereafter \corot) do not
introduce any radial heating and orbits preserve their circularity in this linear
limit.

In a reference frame that is rotating with the spiral perturbation, particles
can be `trapped' at the \corot.
Such particles experience significant oscillations in the radial direction.
For an individual transient spiral perturbation, \tsb find that the
magnitude of angular momentum changes peaks near the \corot and,
therefore, a succession of transient spirals with different pattern speeds can cause
angular momentum changes in the entire stellar disc.
\tsb argue that such spiral perturbations have to be transient because otherwise they
would undo the (radial) changes in stellar orbits they caused
\citep[see also section III.C.1. in the review by][]{Sellwood2014}.

\citet{Roskar2012} find that stars may be transported to the \corot of another
coexisting transient spiral perturbation and therefore may migrate significant distances
in a short time without being heated.

Steady (\ie not transient) rotating perturbations are not expected to modify
stellar orbits that are not
in resonance with the perturbation \citep{Lynden-Bell1972}.
Such resonances occur at \corot (see above) and at \emph{Lindblad resonances}, where the
frequency at which a star experiences the perturbation $m\left(\om\star-\om p\right)$
equals the absolute value of its radial frequency.
$m$ is the azimuthal multiplicity of the perturbation.
This condition can be satisfied inside and outside of the \corot.
The corresponding resonances are being referred to as the \emph{inner} (ILR) and
\emph{outer} Lindblad resonances (\olr).
At the Lindblad resonances, orbits will experience radial heating (equation
\eqref{eq:heating_radial_migration}).
However, the widely accepted paradigm is that \corot are the dominant
source of radial motions due to spiral perturbations \psb.
\tsb also find that particles on near circular orbits migrate more efficiently than those
on eccentric orbits.
\changed{This bias has been confirmed in later works by \citet{Solway2012,Vera-Ciro2014}.
These studies also find a similar bias towards stars that are dynamically cool in the
vertical direction.}
The terms \emph{radial migration} or \emph{churning} have been established for the
above mechanism \citep{Schoenrich2009,Sellwood2014}.
We will use both terms interchangeably throughout the rest of this paper.

It has long been known that bars cause considerable radial mixing.
Migration can also be generated by a bar perturbation but,
unlike churning, bar induced mixing dynamically heats stars \citep{Hohl1971} and gas
\citep{Friedli1994}.
Bar and spiral resonances can overlap and cause chaotic or non-linear evolution of
stellar orbits \citep{Quillen2003}.
Such overlaps may move stars across the disc even more effectively
than churning \citep{Minchev2010,Minchev2011}.
Bars have also been linked to the formation of outer rings
\citep{Romero-Gomez2006,Romero-Gomez2007,Athanassoula2009} indicating that they can drive
radial redistribution.
However, the formation of rings is generally believed to be caused by gas that it driven
to the \olr by bars \citep{Mo2010}.

\tsb point out that despite the significant amount of radial migration in their simulations,
there is no significant influence on the morphology of the stellar disc.
Bars, however, are known to significantly change a disc's appearance \citep{Hohl1971}.

\citet{Roskar2008} showed that churning can move stars beyond the truncation
radius of simulated discs ultimately forming the outer part of a Type-II disc
profile.
The predicted `U-shaped' age profile for this scenario has been confirmed by observations
(see section \ref{sec:explained_breaks}).
However, churning could not be linked to Type-III profiles.

In this paper we will show that bars can, indeed, be responsible for Type-III
disc breaks by substantially heating a subset of disc stars.
We will also present arguments that disfavour churning in \changed{\citet{Roskar2008}
as the cause for the outward migration} occurring in our slowly rotating
simulations from \thfifteen.

\section{Simulations}
\label{sec:simulations}

We set up a $10^{12}\msun$ NFW \citep{NFW} halo with 10 \% of its mass in gas.
The halo concentration is $c=10$.
Dark matter (DM) and initial gas particle masses are $1.1\times10^6\msun$ and
$1.2\times10^5\msun$ respectively.
The gravitational softening length is $\epsilon_\mathrm{soft}=227\,\pc$.

The DM particles' velocities were drawn from the equilibrium distribution function
\citep{Kazantzidis2004}.
The gas particle velocities were set according to the cosmologically motivated angular momentum
profile from \citet{Bullock2001} and are purely tangential with no velocity dispersion as
that is modelled by the hydrodynamics solver.
The velocities are normalized according to the halo's spin parameter $\lambda$ which is
the free parameter in our set-up.
We use the definition from \citet{Bullock2001}:
\begin{equation}\lambda=\left.\frac J{\sqrt{2}MVR}\right|_{R=R_{200}}\end{equation}
The initial temperature profile satisfies hydrostatic equilibrium.

For further details about the initial conditions the reader is referred to
\thfifteen.%
\footnote{The initial conditions code {\sc pyICs} is publicly available and can be
downloaded from \url{https://github.com/jakobherpich/pyICs}.}
Since the simulations are self-consistent, the level of control we have is limited,
\ie we are not able to tune certain features of the simulated galaxies independently,
\eg the bar strength or the disc scale length.

The simulations were evolved with a slightly modified version of the publicly available
treeSPH code {\sc ChaNGa} \citep{Jetley2008, Jetley2010, Menon2015}.
The employed version features stochastic star formation \citep[$c_\star=0.1$]{Stinson2006}
following the Kennicutt-Schmidt law \citep{Kennicutt1998}, radiative metal line cooling,
metal diffusion and stellar feedback \citep{Stinson2013, DallaVecchia2012}.
This recipe has proven to produce realistic galaxies in cosmological simulations
\citep{Wang2015}.

The simulations were evolved for a period of 8 Gyr.
We explored the spin parameter range of $0.02\le\lambda\le0.1$.
For the analysis we used the $N$-body analysis framework {\sc pynbody} \citep{pynbody}.

\section{Results}
\label{sec:results}

\begin{figure*}
\begin{center}
\includegraphics[width=.85\textwidth]{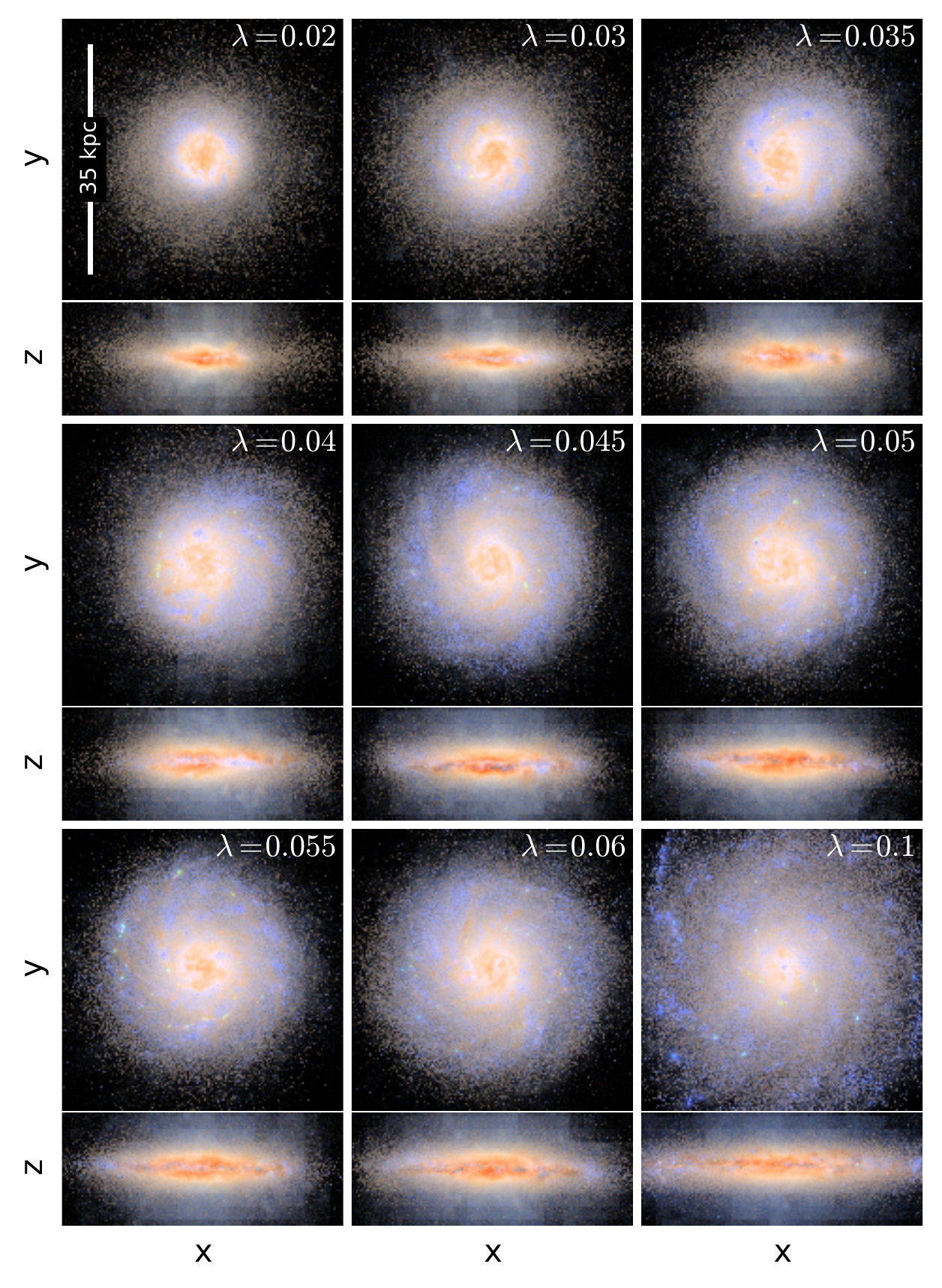}
\end{center}
\caption[Mock observations]{Mock observational images of all simulated galaxies.
    The image shows face-on and edge-on images of all simulated galaxies from low spin
    ($\lambda=0.02$) in the top left to high spin ($\lambda=0.1$) in the bottom right.
    Each panel is $35\,\kpc$ across.
    The images have been created with the radiative transfer code {\sc sunrise}
    \citep{Jonsson2006}.
    The low spin galaxies are compact and red but they do have a disc component which is 
    clearly visible in the edge-on view.
    The edge-on views also reveal a massive central bulge for the lowest spin simulations.
    As the initial spin increases the galaxies get more and more radially extended and less
    centrally concentrated.
    All galaxies with $\lambda>0.04$ show signs of spiral structure in the blue components.
    In all cases the discs are rather thick.
}
\label{fig:sunrise}
\end{figure*}

\begin{figure*}
\includegraphics[width=\textwidth]{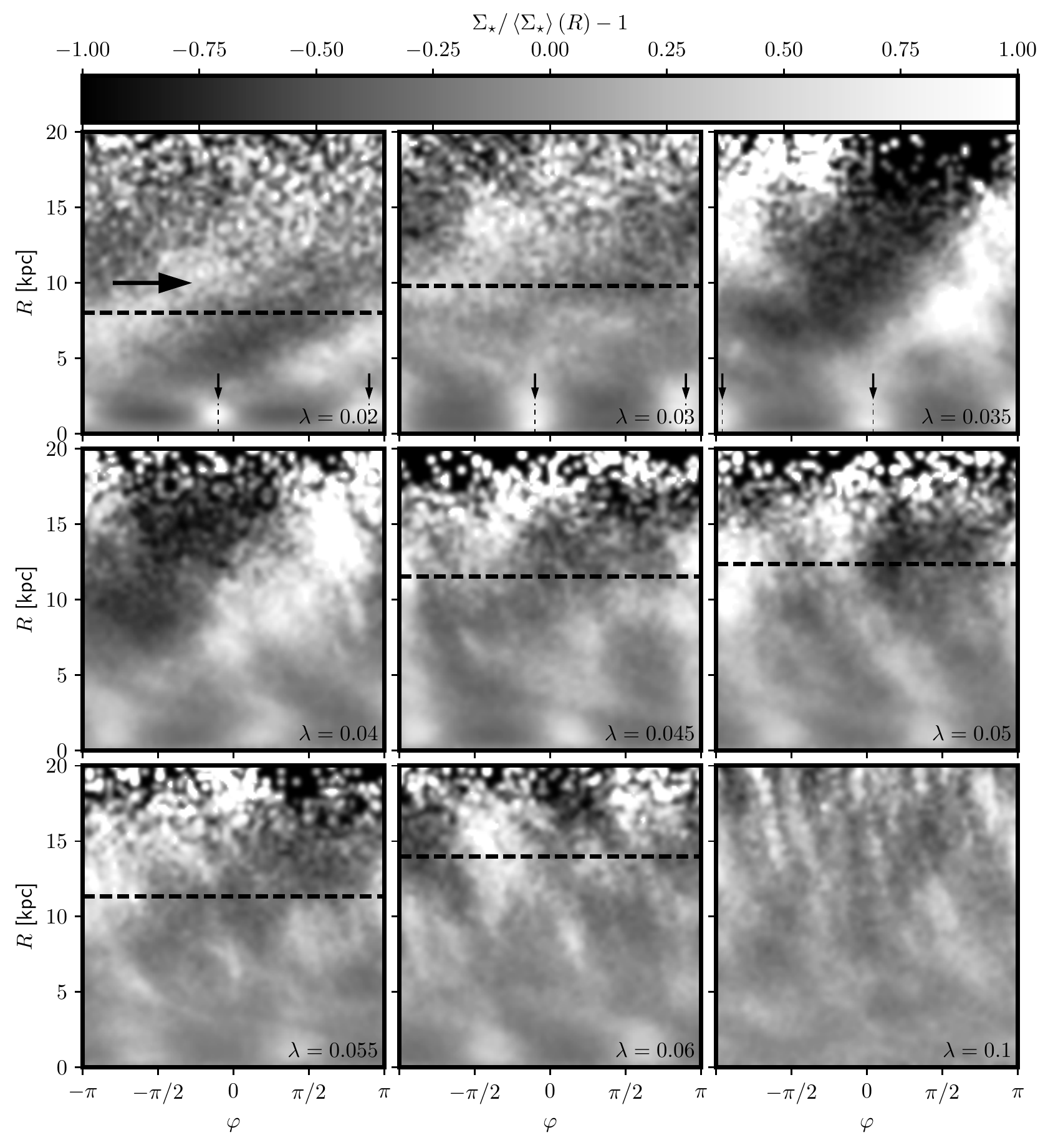}
\caption[Relative stellar density in polar coordinates]{
    
    The plot shows the relative stellar overdensity 
    $\delta=\Sigma_\star\left(R, \varphi\right)/\left<\Sigma_\star\right>\left(R\right)-1$
    in polar coordinates for each simulated galaxy after 8 Gyr.
    The horizontal arrow in the top left panel indicates the sense of rotation of
    all galaxies.
    The horizontal dashed lines indicate the position of the break (except for the
    type-I discs, for $\lambda=0.1$ the break is located just outside the plotted range).
    The small vertical arrows and corresponding dotted lines in the top row indicate the
    position of the bars in the lowest spin simulations.
    
}
\label{fig:polar}
\end{figure*}

In the present paper we study the orbit evolution of individual stellar particles in
the lowest spin simulation from \thfifteen
which forms an Type-III disc break ($\lambda=0.02$).
The goal is to determine how these profiles form.
Throughout the rest of this paper we will use the term \emph{outer disc} to refer to
stars which are located outside the position of the break in the radial profile
in the final simulation output after 8 Gyr.

In Fig. \ref{fig:sunrise} we present mock observational images of all simulated
galaxies from \thfifteen.
They were created using the radiative transfer code {\sc sunrise} \citep{Jonsson2006}.
The galaxies are more extended and less radially concentrated as $\lambda$ increases.
All galaxies are disc-like with albeit rather thick discs.
Fig. 1 in \thfifteen presents the surface density profiles of the respective galaxies
(see \thfifteen for details).

In Fig. \ref{fig:polar} we present the average stellar overdensity relative to the
mean density
$\delta=\Sigma_\star\left(R, \varphi\right)/\left<\Sigma_\star\right>\left(R\right)-1$
at the respective radius in polar coordinates.
For the lowest spin simulation (top left panel) we see a small but strong bar
(vertical features offset by $\Delta\varphi=\pi$ indicated by the small vertical arrows).
The bar signature gets weaker with increasing spin and completely disappears for the highest
spin simulation.
While the discs of the high spin galaxies ($\lambda>0.04$) feature complex trailing patterns,
the lowest spin galaxies exhibit a one-armed leading pattern.
This is consistent with the visual impression from Fig. \ref{fig:sunrise}.

The middle panel of Fig. 2 in \thfifteen shows the ratio of
inner to outer scale lengths as a function of $\lambda$ from a broken exponential fit.
It shows that a high initial spin yields galaxies with Type-II disc surface
density profiles, while low initial spin yields Type-III profiles.

\subsection{Outer discs}
\label{sec:outer_discs}

Based on very similar simulations, \citet{Roskar2008} reproduced the formation of
Type-II disc breaks.
The position of the break in the disc profile is set by a radial cut-off of star
formation due to the radial termination of the gaseous disc.
In their simulations, stars later found in the disc outskirts formed inside the
break radius.
They found no evidence for radial heating causing migration but instead resonant scattering
off of transient spiral arms \psb to be responsible for the outward migration
of stars.
An important signature of this mechanism is that the migrated stars in the outer disc
are predominantly on near-circular orbits, \ie rotationally supported
\citep[see also][]{Roskar2012}.
In this section we will show that the migration mechanism in our high-spin
simulations is consistent with this mechanism.
These can produce Type-II disc profiles as well.
We also show that the low-spin simulations, which exhibit Type-III breaks, are not.

We first confirm the finding of \thfifteen that all outer disc stars in all of the
simulated galaxies must have migrated outward from inside the respective break radii.
Additionally, we will show that the outer disc stars in the discs with Type-III
profiles stem from the vicinity of the galactic centre, while their counterparts
in the discs with Type-II profiles were born further out in the disc.
This will be followed by showing that the outer disc stars in the Type-III cases
are on very eccentric orbits.

\begin{figure}
\includegraphics[width=\columnwidth]{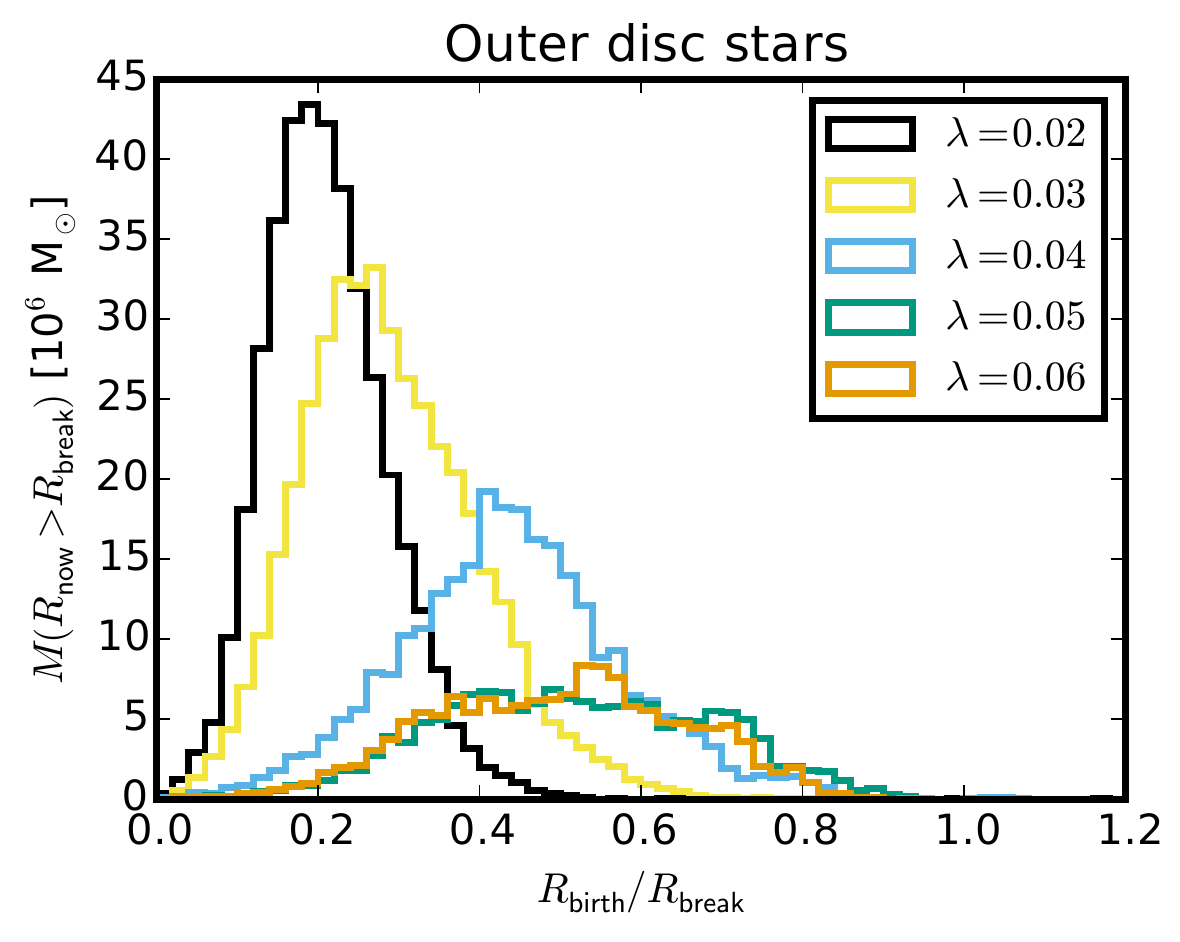}
\caption[Distribution of birth radii for stars in outskirts]{The distribution of birth radii
    for outer disc stars.
    The radii are normalised by the break radius in the final snapshot.
    Practically all stars beyond the break region were born at smaller radii.
    The effect is most extreme for the Type-III breaks ($\lambda\lta0.03$).
    }
\label{fig:rxy_hist}
\end{figure}

Fig. \ref{fig:rxy_hist} shows the distribution of birth radii of the stars in the outer disc.
We see that practically all outer disc stars are born inside the respective break
radius.
Here, we find a stark difference between the low- and high-spin cases.
In low-spin simulations almost all outer disc stars are born well inside the
break radius, $R_\mathrm{birth}<R_\mathrm{break}/3$.
In the high-spin cases the birth radii are located at larger galactocentric distances,
$R_\mathrm{birth}\approx R_\mathrm{break}/2$.
This difference is even more clear in absolute values as the break radii grow with $\lambda$
\phfifteen.
We summarize that in all cases the outer disc stars migrate from inside the break region and
in the case of the lowest spin simulations originate from predominantly very small radii.

We now characterize stellar orbits in our simulations through their circularity
parameter $\jzjc$.
The circularity parameter is the $z$-component of a star's
specific angular momentum normalized by the specific angular momentum of a circular orbit
at the same specific orbital energy.
By construction a particle on a perfectly circular orbit has $\jzjc=1$.
Particles on perfectly radial orbits satisfy $j_z=0$ and thus have a circularity parameter of $0$.
In practice we will consider every particle with $0.8\lta\jzjc\lta1$ to be on a
near circular orbit 
and particles with $\jzjc\approx0$ on radial orbits.
Eccentric orbits cover the range in between these two limiting cases.
Negative values correspond to counter-rotating orbits.

\begin{figure}
\includegraphics[width=\columnwidth]{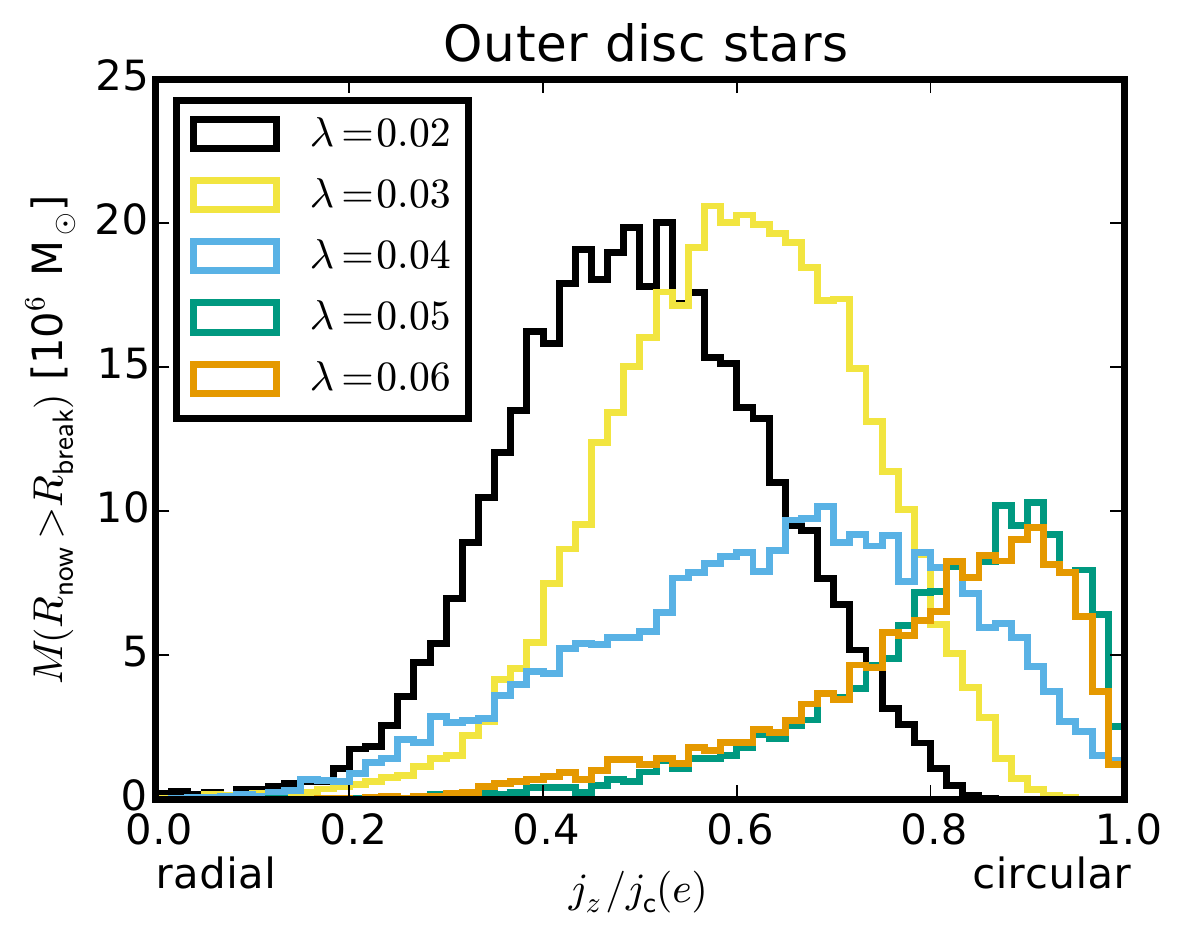}
\caption[Circularity distribution outer disc]{Circularity distribution in outer disc.
    The figure shows the circularity distribution of stars in the disc outskirts ($R>\rb$)
    for a choice of simulations spanning the
    range from low ($\lambda=0.02$) to high spins ($\lambda=0.06$).
    Most orbits in the high spin simulations ($\lambda > 0.04$) are circular while there is
    not a single star with $\jzjc\gta0.85$ in the lowest spin simulation.
    Intermediate spin simulations show a broad circularity distribution in the outskirts
    ranging from completely radial ($0$) to circular ($1$).
    For pure exponential discs we interpolated the ``break radius''.
}
\label{fig:jzjc_hist_outskirts}
\end{figure}

In Fig. \ref{fig:jzjc_hist_outskirts} we present mass-weighted histograms of the distribution
of \jzjc for stars which are beyond the break radius for selected simulations.
For the Type-II breaks ($\lambda>0.04$) most of the orbits are near
circular and thus consistent with the mechanism proposed by \citet{Roskar2008}.
The presence of complex spiral patterns in those galaxies (see Fig. \ref{fig:polar})
supports this hypothesis.
The simulations with Type-III breaks, on the other hand, only have very few
stars in their
outskirts that are on near circular orbits.
Contrary to the higher spin simulations, the distribution of the circularity parameter
drops to $0$ at values of $\approx0.8-0.9$ in these cases.
The \jzjc distribution peaks roughly at $0.5$ which corresponds to
rather eccentric orbits.
This lack of near circular orbits beyond the break in our simulations with
Type-III disc breaks
disfavours radial migration as described in section \ref{sec:dynamics} as a possible
formation mechanism because it is less efficient for eccentric orbits \psb.

As there is no evidence that the formation mechanism of the Type-II breaks
differs from that of \citet{Roskar2008}, we will focus our further analysis on the yet
unexplained Type-III breaks.

\subsection{Orbit evolution in Type-III discs}
\label{sec:star_orbit}

Given the substantial radial mass redistribution in Type-III discs,
we are now going to explore what mechanism is most important in driving this
redistribution.
We focus our analysis on the lowest spin simulation ($\lambda=0.02$).
The break radius of this galaxy is $R_\mathrm{break}=8.0\pm0.5\,\mathrm{kpc}$.

\begin{figure}
\includegraphics[width=\columnwidth]{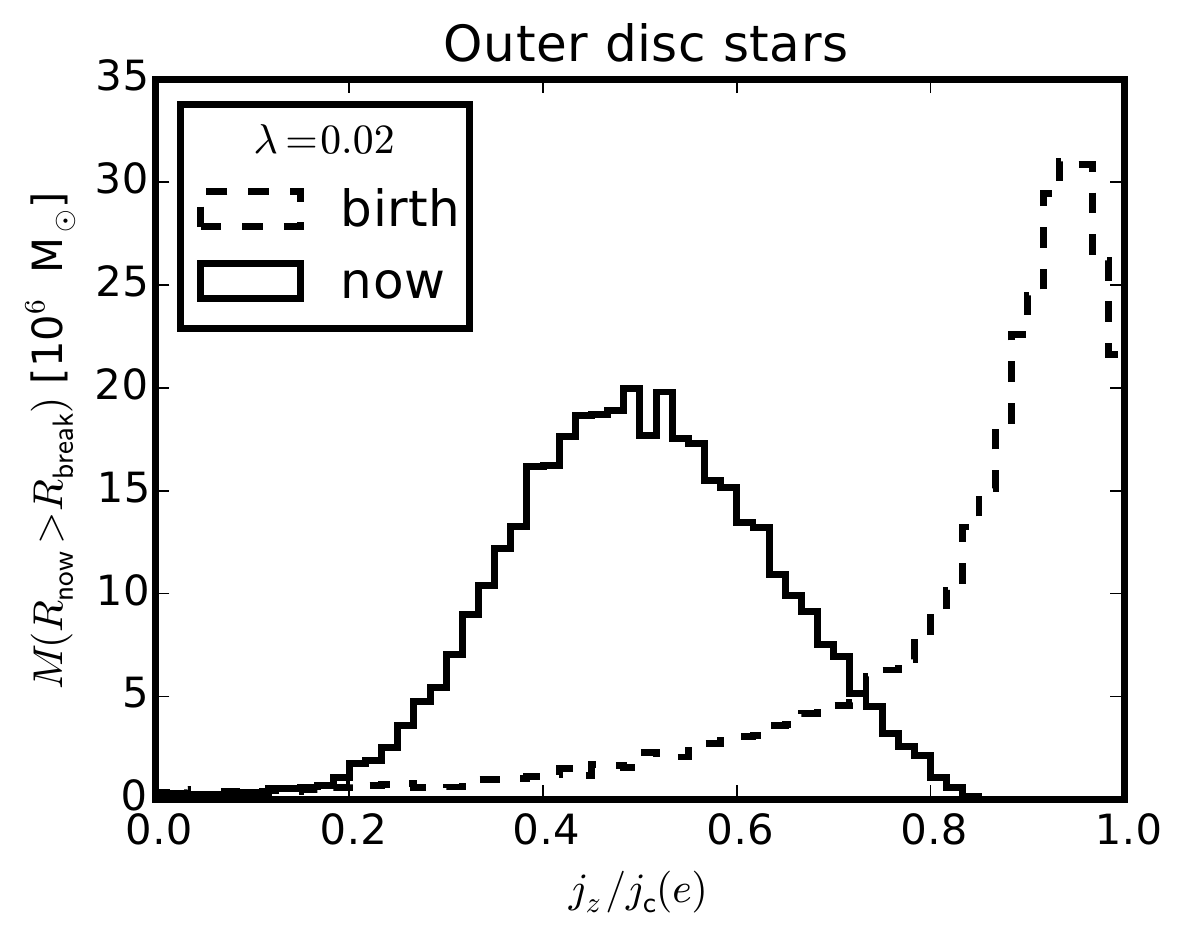}
\caption[Circularity evolution]{Circularity evolution in a low spin simulation.
    The figure shows the distribution of the circularity parameter \jzjc of the stars
    outside the break radius at
    their formation (dashed) and in the final snapshot (solid).
    When these stars are born, most of them are on a circular orbit but as they migrate
    outwards, their orbits get more and more eccentric.
}
\label{fig:jzjc_hist_ev}
\end{figure}

Fig. \ref{fig:jzjc_hist_ev} shows the distribution of the circularity parameter \jzjc of the
outer disc compared to the star's \jzjc distribution at birth (dashed line).
Most of the outer disc stars were born on nearly circular orbits.
Thus, there must be a mechanism which transforms orbits from circular to radial and
significantly increases the orbits apocentre.

\begin{figure}
\includegraphics[width=\columnwidth]{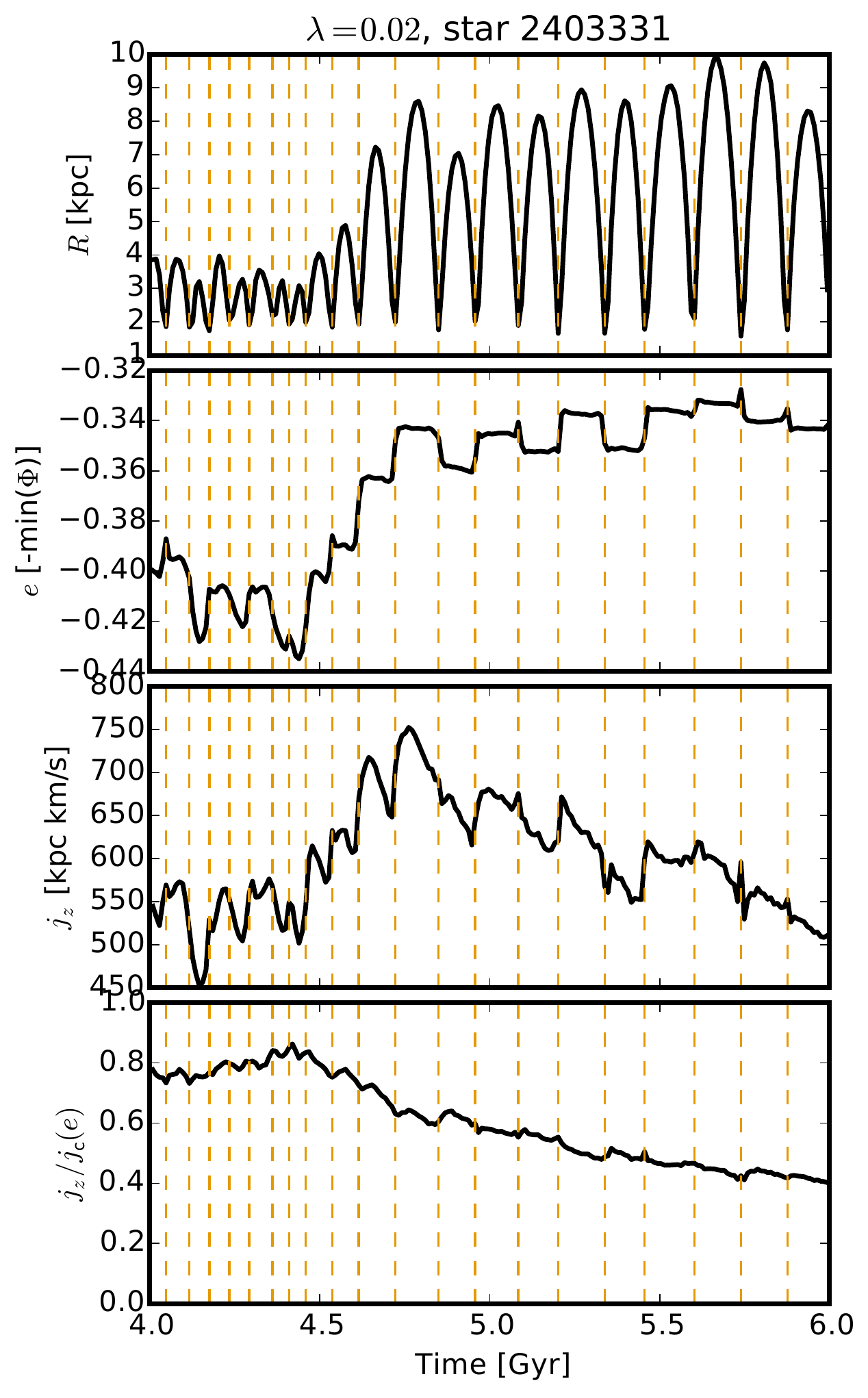}
\caption[Stellar orbit evolution]{Orbit evolution of an individual star.
    The vertical dashed lines indicate the pericentre passages of that star.
    The top panel shows the evolution of its radial position, the second panel shows
    its orbital energy, the third panel
    shows the evolution of its $z$-component of specific angular momentum and the
    bottom panel shows its circularity parameter.
    {
    While the star's apocentre is still small ($t\lta4.5\,\mathrm{Gyr}$) its binding
    energy is almost permanently changing.
    After approximately 4.5 Gyr its apocentre growths significantly and the star's binding
    energy turns into a step function with steps occurring only at pericentres.
    The star particle instantaneously gains energy and angular momentum during many
    pericentre passages.
    In between these passages its energy stays constant but it looses angular momentum which
    causes the particles circularity to decrease steadily.
    }
}
\label{fig:star_orbit}
\end{figure}

In order to figure out how orbits get transformed we looked into the time evolution of the
orbit of one particular stellar particle.
Fig. \ref{fig:star_orbit} presents the evolution of orbital parameters of a sample star
particle from the outer disc.
The four panels show (from top to bottom) the evolution of the star's radial position $R$,
specific orbital energy $e$, specific angular momentum $z$-component $j_z$
and circularity parameter \jzjc.
The times of pericentric passage are marked by vertical dashed lines.

After 4.5 Gyr the star gains or loses energy (second panel) and angular momentum
(third panel) at each pericentre passage.
When the star is far away from pericentre, its energy stays constant,
but it loses specific
angular momentum (third panel), which reduces the circularity of its orbit (bottom panel).

During this time span, the binding energy evolves approximately following a step function.
The steps occur at pericentre.
If the star gains energy $e$ at pericentre its semi-major axis increases and vice versa.
The pericentre, however, stays more or less constant at approximately 2 kpc.
This particular star has a net gain in energy which is not surprising as it has been
selected from the outer disc in the final simulation output.
In general it is also possible for a star to have a net loss of energy.
Such stars migrate towards the galaxy centre and, therefore, cannot be found in the outer disc.

Before 4.5 Gyr, the star in Fig. \ref{fig:star_orbit}
has a rather small semi-major axis ($<4\,\mathrm{kpc}$) and the steps in
the evolution of its binding energy are smoother and not as well defined yet.
Generally, the binding energy of the star is constant when it is far away from the
centre and varies only near the centre.
Thus, we expect the source of the variation of binding energy to be in the centre of the galaxy.
As we will show later (sections \ref{sec:bar_star} and \ref{sec:toy_model}) the source is
the central bar.
We stress that the behaviour of the star displayed in Fig. \ref{fig:star_orbit} is
qualitatively similar to that of other stars found in the outskirts of our low
spin discs.

\subsection{The bar in Type-III discs}
\label{sec:bar}

Since particle orbits seem most affected in the centre of galaxies near a possible bar.
We characterize the bar using the coefficient of the $m=2$ Fourier mode:
\begin{equation}
A_2 = \sum_j \exp\left(i2\varphi_j\right) m_j
\end{equation}
where $m_j$ and $\varphi_j$ are the mass and azimuthal angle of the stars.
The sum is over all stars in the considered region\footnote{\eg annular bins or spherical
regions of variable size} of the galaxy.
This $m=2$ mode encodes two relevant quantities, namely the relative bar strength
\begin{equation}
A_2/A_0 = \frac{\left|A_2\right|}{\sum_jm_j}
\end{equation}
and the bar's position angle
\begin{equation}
\theta_2 = \frac{\arg\left(A_2\right)}2\mod\frac{2\pi}2
\label{eq:bar_angle}
\end{equation}

\begin{figure}
\includegraphics[width=\columnwidth]{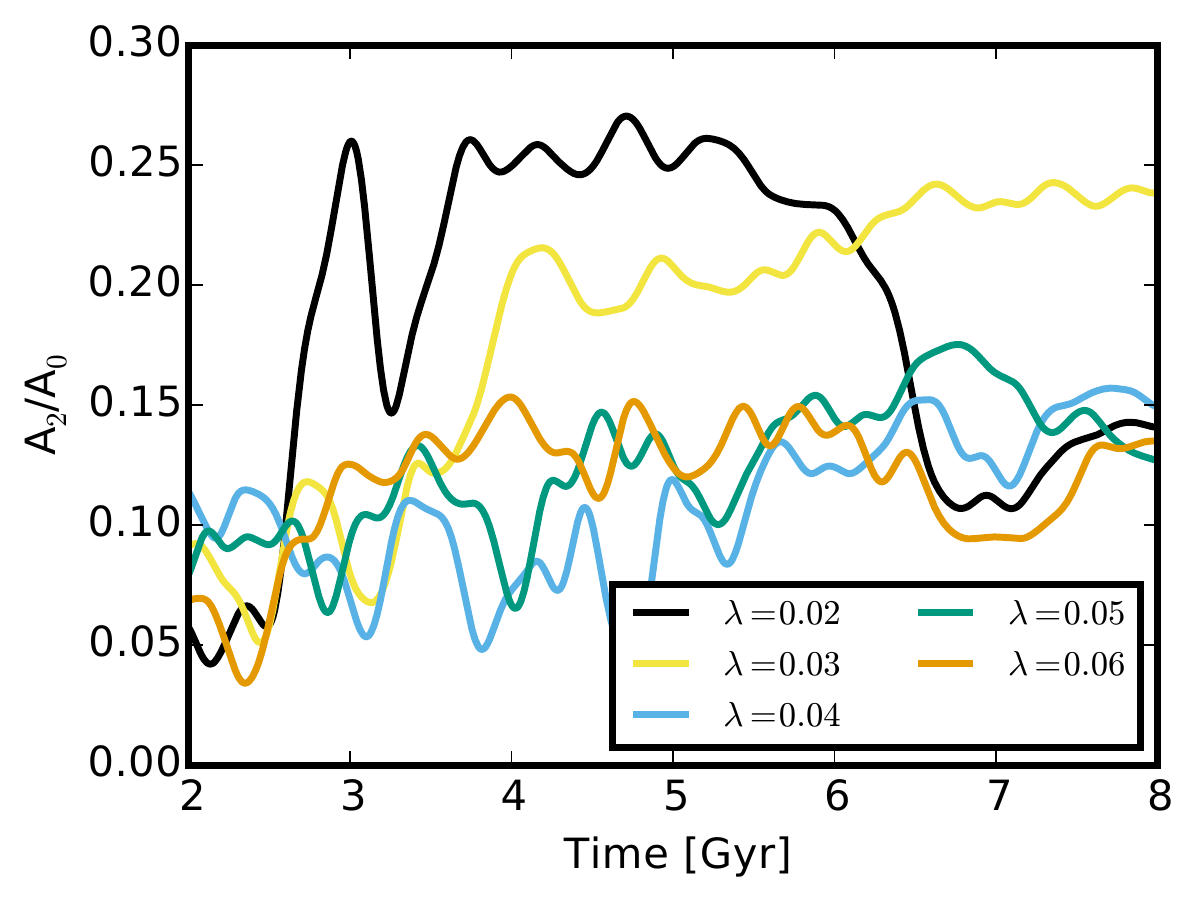}
\caption[Bar strength evolution]{Bar strength evolution.
    The figure shows the relative amplitude of the $m=2$ Fourier mode of all stellar particles
    inside 1 kpc as a function of time.
    While the values for intermediate and high spin are rather low, there is an excess in
    bar strength for the lowest spin ($\lambda=0.02 - 0.03$) simulations for a period of
    about 4 Gyr.
}
\label{fig:bar_strength}
\end{figure}

In Fig. \ref{fig:bar_strength} we present the time evolution of the bar strength
in the innermost kpc for a selection of simulations.
The lowest spin simulation has the strongest
bar.
Its bar strength exceeds that of the high spin simulations for a continuous period of
4 or more Gyr.
During this period its value is remarkably constant.

The only other simulation that exhibits a strong bar is the one with the second lowest
initial spin.
Thus, the only simulations in our sample with a strong bar are also the only ones
which form a Type-III break.
This correlation encouraged us to search for a possible dynamical link between bars and
Type-III breaks in our simulations.

\begin{figure}
\includegraphics[width=\columnwidth]{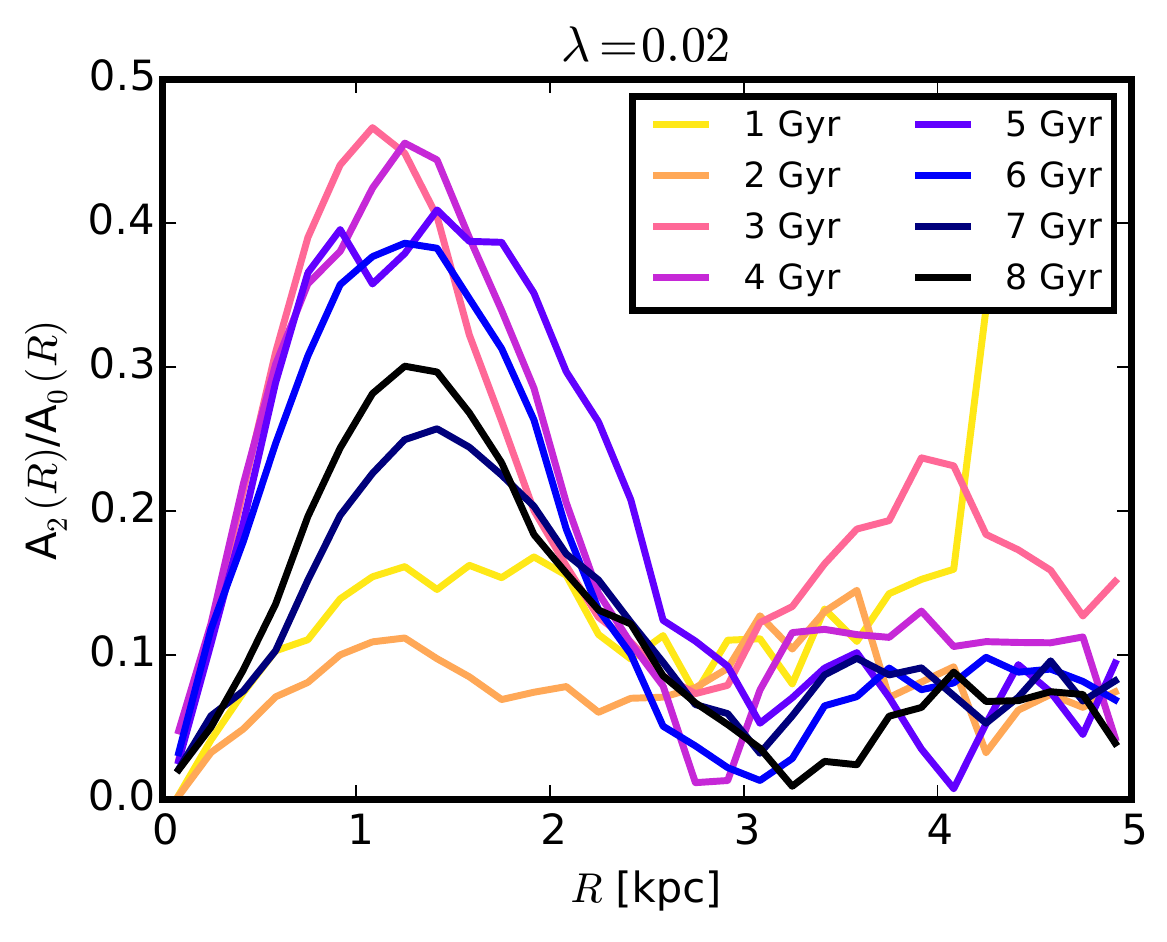}
\caption[Local bar strength evolution]{Evolution of the bar strength profile.
    The plot shows the relative amplitude of the $m=2$ mode in radial bins
    of the stars in the simulation at different times.
    We see that the bar forms after 2-3 Gyr (see also Fig. \ref{fig:bar_strength})
    and that it extends out to 2 kpc.
    }
\label{fig:bar_profile_ev}
\end{figure}

Fig. \ref{fig:bar_profile_ev} shows the relative bar strength in different radial bins
at different output times for the $\lambda=0.02$ galaxy.
The size of the bar does not evolve significantly.
It ends at $2-3\,\mathrm{kpc}$, so it makes sense to look for possible effects on star particles
inside that region.

\begin{figure}
\includegraphics[width=\columnwidth]{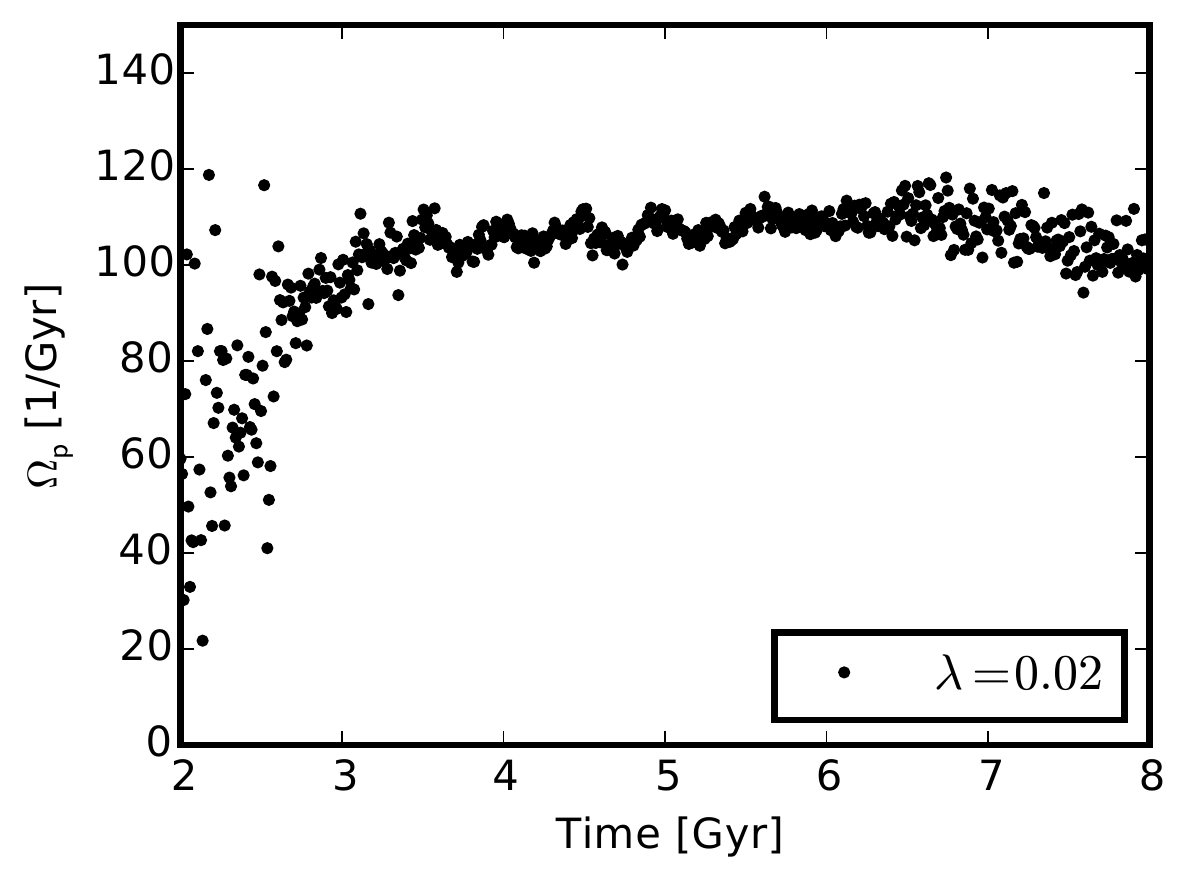}
\caption[Bar pattern speed]{Bar pattern speed evolution in the low spin simulation.
    The figure shows the bar pattern speed of the $\lambda=0.02$ simulation as a function of time.
    Before its formation ($\approx 3\,\gyr$, see Fig. \ref{fig:bar_strength}) it is not well
    defined but later on it is constant.
}
\label{fig:pattern_speed}
\end{figure}

The pattern speed of the bar is the time derivative of the bar's position angle.
Fig. \ref{fig:pattern_speed} presents the pattern speed of the bar in the low
spin ($\lambda=0.02$) simulation as a function of time.
Once the bar has formed, its pattern speed is remarkably constant, which
will be important when we discuss the interaction of
stellar particles and the bar in section \ref{sec:bar_star}.

\begin{figure*}
\includegraphics[width=\textwidth]{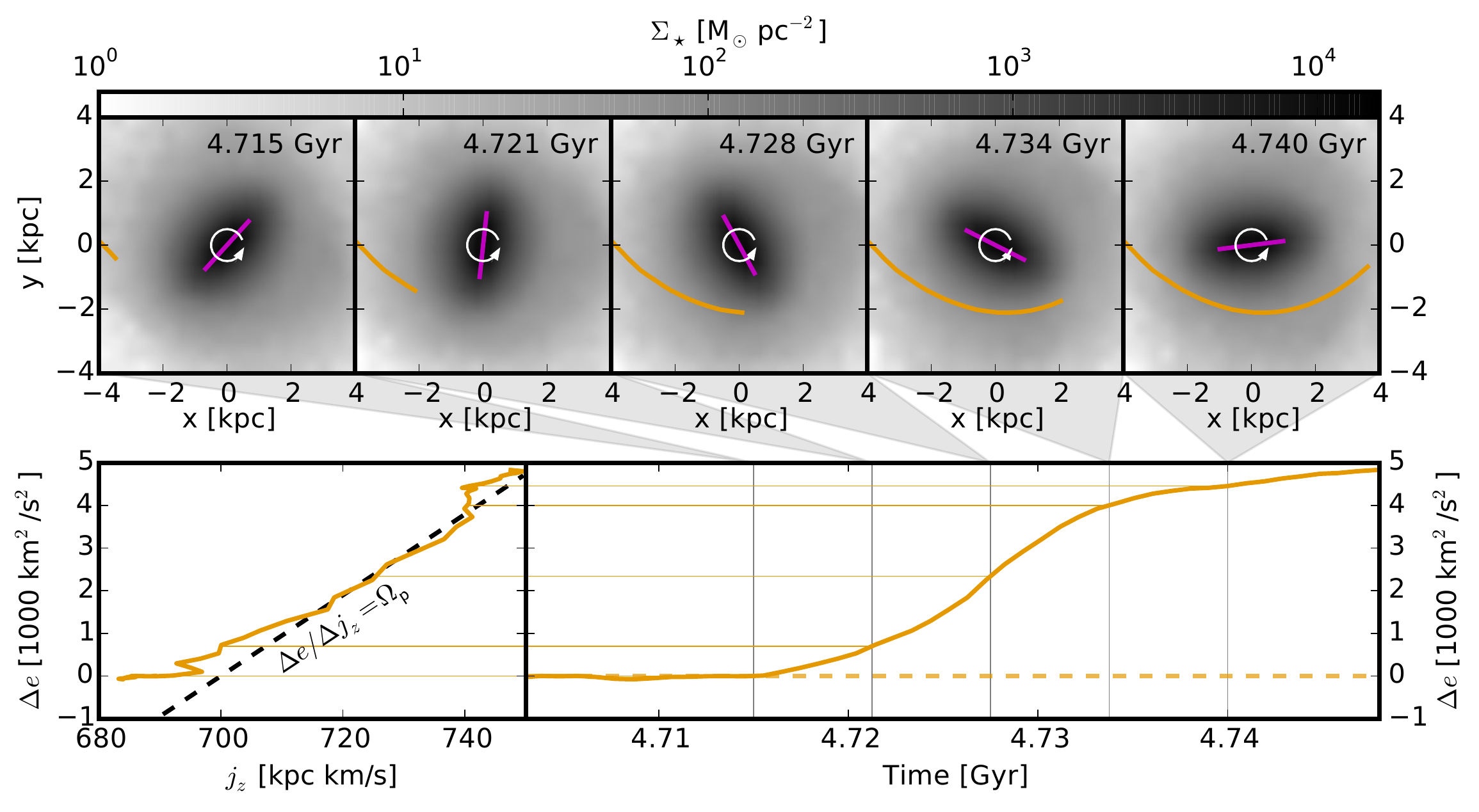}
\caption[Phase comparison]{%
    The panels in the top row
    show the trajectory of an individual star in the low spin simulation
    during one pericentre passage.
    The stellar surface density is projected as a greyscale in logarithmic scale.
    The bar's position is indicated by the magenta line.
    Its sense of rotation is indicated by the white arrow.
    The star is the same as in Fig. \ref{fig:star_orbit}.
    The bottom left panel shows the star's trajectory in the Lindblad diagram.
    The slope of the dashed line in the bottom left panel is the bar's pattern speed \om{p}
    indicating the trajectories of constant Jacobi energy.
    The bottom right panel shows its relative binding energy $e$ as a function of
    time.
    The binding energy is given relative to its value before the encounter which is indicated
    by the orange dashed line.
    The corresponding time of each panel in the top row is indicated by a vertical line in
    the bottom right panel.
    
    The figure shows that significant evolution of the star's orbital energy occurs during
    pericentre passage.
    Its trajectory in the Lindblad diagram is consistent with a constant Jacobi energy
    during pericentre.
    This indicates that the bar-induced potential perturbation is responsible for the
    energy gain.
}
\label{fig:bar_star}
\end{figure*}

To summarize, low angular momentum simulations develop strong
and long-lived bars that have a stable pattern speed over a number of Gyrs.
When the bar forms it may trap particles at its resonances.
These stars are expected to oscillate radially \psb but stay bound to the bar.
Note, that such strong bars are known to cause considerable heating of stellar orbits
\citep{Hohl1971}.

\subsection{Orbit migration from star-bar interaction}
\label{sec:bar_star}

In the top row of
Fig. \ref{fig:bar_star} we show the current position of the star particle
from Fig.
\ref{fig:star_orbit} during one direct interaction with the bar that significantly
increased its binding energy $e$ and apocentre.

The panels in the top row
of the figure show that the star and the bar have the same sense of
rotation and that the azimuthal phase of the star lags behind that of the bar.
During the period that is shown in these images, the star gains a significant amount of
binding energy $e$ (bottom right panel).
The slope of the star's trajectory in the Lindblad diagram (bottom left panel) is
remarkably similar to \om p, particularly for $4.721<t/1\,\mathrm{Gyr}<4.734$ which is the
time period in which the time evolution of the binding energy $e$ is the strongest.
According to equation \eqref{eq:delta_e},
this is the trajectory we expect from the interaction of a star with the bar,
given its constant pattern speed.
Taking into account that the orbital binding energy only changes in the vicinity of the bar
(see Fig. \ref{fig:star_orbit}),
we interpret this as a clear signature of the bar as the dominant driver of the star's
energy gains (losses).

The encounter of the star and the bar are analogous to a \emph{swing-by}.
The star is exposed to an extended period of acceleration due to the moving potential
well of the bar---it is \emph{surfing} in the bar's potential.

So far we investigated the behaviour of a single star.
However, finding a mechanism that drives a single star's migration out into the outer disc
does not explain the existence of the outer disc.
In order to show that the behaviour of that star is typical, we will show
that there is a prominent signature of the bar in the outer disc stars' orbit evolution.%
\footnote{This signature can also be caused by different non-axisymmetric perturbations
that are rotating with the same pattern speed.
However, the only such perturbation that was found in the simulation is the bar.
}
We will use equation \eqref{eq:delta_e} which implies that the ratio of $\Delta e$ and
$\Delta j_z$ equals the bar's pattern speed \om p.

\begin{figure}
    \includegraphics[width=\columnwidth]{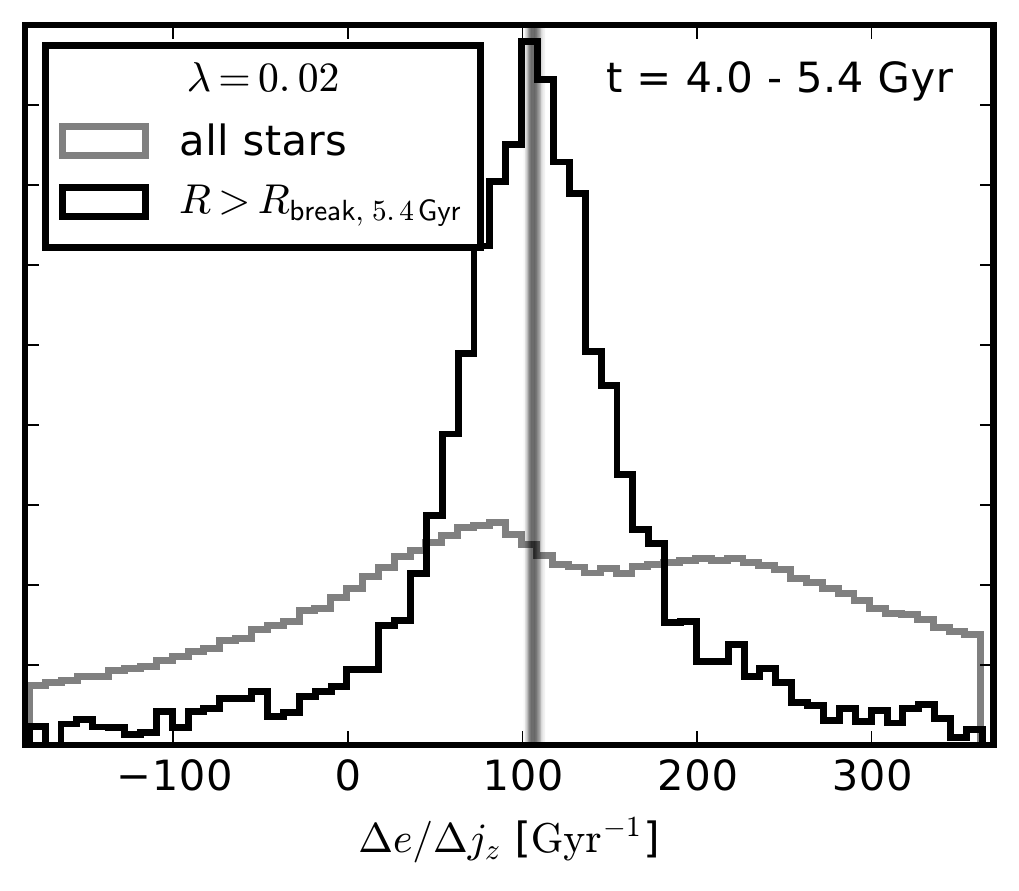}
    \caption{
        Mass weighted distribution of $\Delta e/\Delta j_z$ of stars located outside
        the break radius at 5.4 Gyr
        (black) and all stars (grey) of the lowest-spin ($\lambda=0.02$) simulation.
        The time range (4.0 - 5.4 Gyr) was chosen to coincide with the times where the bar
        strength is at its
        peak value and its pattern speed is approximately constant simultaneously
        (see Figures \ref{fig:bar_strength} and \ref{fig:pattern_speed}).
        The vertical lines indicate the measured values of the bar pattern speed that occurred
        in the respective time span.
        It coincides remarkably well with the peak of the distribution for the outer disc
        stars.
        The peak is not present in the global distribution.
    }
    \label{fig:de_dj_hist}
\end{figure}

 Fig. \ref{fig:de_dj_hist} shows the normalized mass-weighted distribution of
$\Delta e/\Delta j_z$ of stars in the $\lambda=0.02$ simulation.
It covers the time range from 4.0 to 5.4 Gyr when the bar was both, strong and had a
constant pattern speed.
The black histogram considers stars that are located outside the break radius at 5.4 Gyr into
the simulation.
The break radius at this time was $R_\mathrm{break,\,5.4\,Gyr}=7.6\,\mathrm{kpc}$.
The grey histogram shows the same for all stars in the simulation that were born before 4.0 Gyr.
Stars now outside the profile break show a clear peak in their orbit evolution of
$\Delta e/\Delta j_z$ which coincides remarkably well with the
bar's pattern speed \om p (vertical lines).
There is a small feature in the global distribution (grey) but it is offset from \om p and
much less prominent.
Given the constant pattern speed of the bar, this agreement for the stars in the outskirts
matches the
prediction of equation \eqref{eq:delta_e} for the imprint of the bar on stellar orbits.
Thus, it suggests that
the bar is the dominant driver of the outward migration happening in the simulations.

\section{`Surfing the bar' in a toy model}
\label{sec:toy_model}

\subsection{Motivation for a toy model}
\label{sec:motivation}
We created a very simple toy model to illustrate that our proposed mechanism
(see section \ref{sec:bar_star}) is indeed capable of increasing the semi-major axes of
stellar orbits.
Here, the only asymmetry will be a bar-like, rotating, non-axisymmetric perturbation in an
otherwise axisymmetric potential.
Its purpose is to support our interpretation from section \ref{sec:results}.
We will show that the interactions of test particles with a bar-like
perturbation can reproduce the qualitative behaviour of stellar particles in the outer
disc of the simulation.
Thus, other effects are not necessary to qualitatively reproduce the kinematics of outer
disc stars.
This supports our result that the bar is the main driver in the formation of Type-III
disc profiles.

\subsection{The time-dependent potential}
\label{sec:toy_model_potential}
In this model, test particles move in a NFW potential
\begin{equation}
\Phi_0\left(\left|\mathbf{r}\right|\right) = \Phi_\mathrm{min}
    \frac{\ln{\left(1+\left|\mathbf{r}\right|/r_s\right)}}{\left|\mathbf{r}\right|/r_s}
\label{eq:nfw}
\end{equation}
with a rotating non-axisymmetric Gaussian perturbation
\begin{equation}
\Phi_\mathrm{p}\left(\mathbf{r}, t\right) =
    f\Phi_\mathrm{min}
    \exp\left[-0.5\left(\frac{{x'}\left(t\right)^2}{\sigma_x^2}+\frac{{y'}\left(t\right)^2}{\sigma_y^2}\right)\right]
\label{eq:pot_pert}
\end{equation}
in the centre, where
\begin{equation}
\begin{pmatrix}x'\\y'\end{pmatrix}\left(t\right) =
    R\left(-\om pt-\varphi_0\right)\begin{pmatrix}x\\y\end{pmatrix}.
\label{eq:rotation}
\end{equation}
In the above equations $\Phi_\mathrm{min}$ is the minimum unperturbed potential, $r_s$ the
scale radius of the NFW potential, $f$ the relative strength of the perturbation,
$\sigma_{x,y}$ the extent of the perturbation in $x$ and $y$ direction%
\footnote{Note that non-axisymmertry requires $\sigma_x\ne\sigma_y$}, $R\left(\varphi\right)$
the rotation matrix for rotation by an angle $\varphi$,
$\om p$ the perturbation's pattern speed and $\varphi_0$ the initial phase of the perturbation's
position angle.
The total potential then is
\begin{equation}
\Phi\left(\mathbf{r}, t\right) = \Phi_0\left(\left|\mathbf{r}\right|\right) +
    \Phi_\mathrm{p}\left(\mathbf{r}, t\right)
\label{eq:total_pot}
\end{equation}
This yields a static NFW potential superimposed with a 2D-Gaussian perturbation that
is rotating with an angular speed of $\om p$ in the positive sense of rotation.
Note, that the particles in this model are \emph{not} self-gravitating.

\subsection{Mass redistribution in the toy model}
\label{sec:toy_model_parameters}
In this section we describe technical details of our toy model that are necessary for
reproducing the results.

We apply this potential to an initially exponential surface density profile with two
different perturbation strengths: $f=0.01, 0.05$.
We refer to them as the \emph{weak} and \emph{strong} model respectively.
The values of the model parameters are given in table \ref{tab:toy_model_pars}.
The initial velocities have been set such that the orbits would be circular in the
azimuthally averaged potential.

We evolved the test particles' coordinates with a Leap-Frog integration scheme.
We used a time step of $10^{-3}\,\mathrm{kpc\,s/km} \approx 10^{-3}\,\mathrm{Gyr}$.
The relative error of the Jacobi-integral is of order $\mathcal{O}(10^{-4})$ except for
a few particles with very small initial radii ($\lta0.02\,\mathrm{kpc}$).
We restricted the treatment to the $x-y$ plane for simplicity.

\changed{In principle this setup is overly simplistic.
We model a disk with particles on perfectly circular orbits and expose it to a bar perturbation
that grows instantly.
This is problematic for two reasons: (i) Circular orbits experience the stronger interactions
with non-axisymmetric perturbations than (mildly) eccentric orbits and hence, the amount of
migration is likely to be overestimated in our simplistic setup
\citep{Sellwood2002,Solway2012,Vera-Ciro2014};
(ii) the instant growth of the bar causes the equilibrium system to be in a non-equlibrium
state which will have to equilibrate again.}

\changed{We performed tests about the effects of both of these problems.
(i) We added a velocity perturbation on top of the initially circular orbit which has a random
direction and the magnitude of the perturbation is drawn from a normal distribution with
a standard deviation of 5\,\%\ of the value of the circular velocity of the respective particle.
(ii) We linearly grew $f$ in time from 0 to its final value with different growth rates.
The duration of the bar-growth phase ranges from 0.5 to 5 Gyr.
Neither of these tests had any measurable effect on the results of the model.
}

\begin{table}

\caption{
    Toy model parameters. $N$ is the number of test particles and $\epsilon_\mathrm{init}$
    their initial orbit eccentricity.
    $R_\mathrm{exp}$ is the scale length of the inital exponential surface density profile.
    All other quantities correspond to those in equations \eqref{eq:nfw}-\eqref{eq:rotation}.
    }
\begin{tabular}{lccc}
\toprule
& Model & weak & strong \\
\cmidrule{3-4}
& Units & & \\
\midrule
$f$ & & $0.01$ & $0.05$ \\
\cmidrule{3-4}
$\Phi_\mathrm{min}$ & [km$^2$/s$^2$] & \multicolumn{2}{c}{$-3\times10^{5}$} \\
$r_s$ & [kpc] & \multicolumn{2}{c}{$10$} \\
$\sigma_x$ & [kpc] & \multicolumn{2}{c}{$2$} \\
$\sigma_y$ & [kpc] & \multicolumn{2}{c}{$\sqrt{2}$} \\
\om p & [rad Gyr$^{-1}$] & \multicolumn{2}{c}{$100$} \\
$\varphi_0$ & [rad] & \multicolumn{2}{c}{random} \\
$R_\mathrm{exp, init}$ & [kpc] & \multicolumn{2}{c}{$1$} \\
$\epsilon_\mathrm{init}$ &  & \multicolumn{2}{c}{$0$} \\
$N$ &  & \multicolumn{2}{c}{$10^5$} \\
$t_\mathrm{final}$ & [Gyr] & \multicolumn{2}{c}{$4$} \\
\bottomrule
\end{tabular}

\label{tab:toy_model_pars}
\end{table}

\begin{figure}
    \includegraphics[width=\columnwidth]{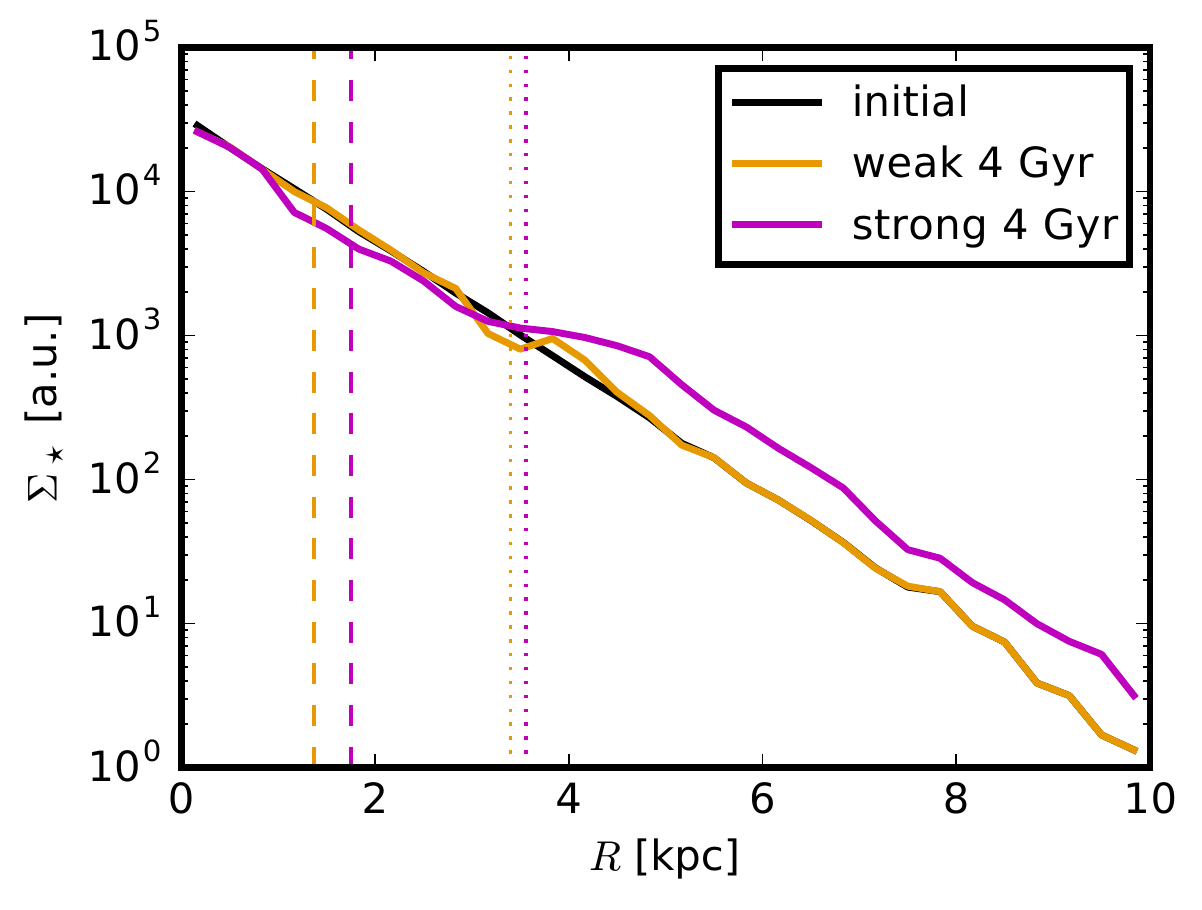}
    \caption{
        The surface density profiles after 4 Gyr for the weak (orange) and strong (magenta)
        models compared to the initial profiles (black).
        Where the black line is not visible, it is hidden behind the orange line.
        The dashed and dotted vertical lines represent the location of the \corot and \olr
        respectively.
        While the strong model shows a dip at \corot and an excess of particles outside of
        the \olr, the profile of
        the weak model remains mostly unchanged except for a small wiggle at the \olr.
    }
    \label{fig:toy_model_profile_ev}
\end{figure}

In Fig. \ref{fig:toy_model_profile_ev} we compare the evolved surface density profile of the
weak (orange) and the strong (magenta) model to the initial (black) profile.
In the weak case there is no significant evolution of the surface density profile.
There is merely a small feature at the location of the \olr (3.4 kpc).
This is consistent with the results of \tsb.

\begin{figure}
    \includegraphics[width=\columnwidth]{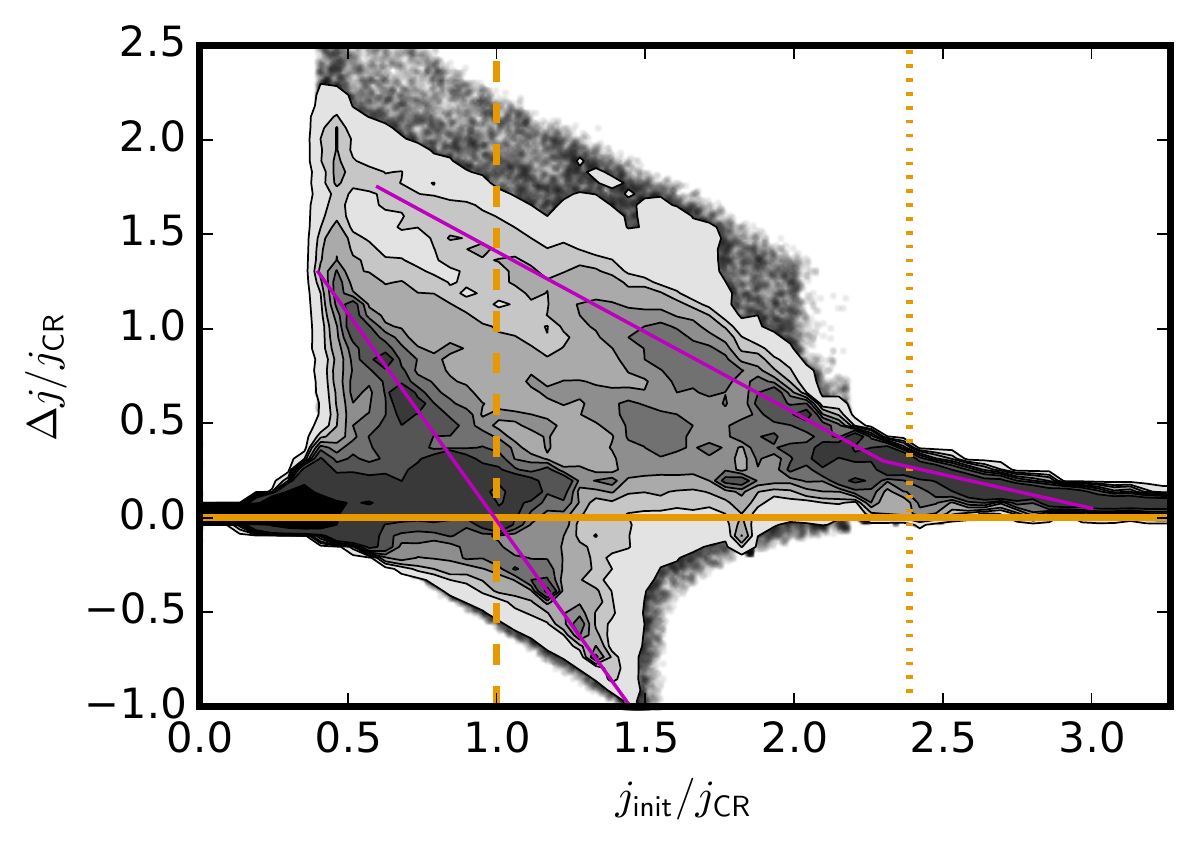}
    \caption{
        The change in specific angular momentum $j$ of the test particles in the strong
        model after 4 Gyr as a
        function of their initial specific angular momentum.
        The horizontal solid line indicates the locus of zero change in $j$.
        It separates outward migration ($\Delta j > 0$) from inward migration ($\Delta j < 0$).
        The dashed and dotted vertical lines indicate the values of $j$ that correspond
        to the \corot and \olr respectively.
        The two most prominent features are highlighted by the magenta lines.
    }
    \label{fig:change_ang_mom}
\end{figure}

The profile of the strong model has a dip at the position of the \corot and just inside the
\olr as well as an excess outside the \olr. 
This indicates that there is significant outward migration from \corot and \olr.
Note that the determination of the position of the \corot and \olr is somewhat dubious
because it is based on the existence of nearly circular orbits.
In a strict sense, such orbits cannot exist in a non-axisymmetric potential.
When we calculate the location of the resonances, we use the azimuthally averaged
potential.
We show later (Fig. \ref{fig:change_ang_mom}) that the resulting resonances agree well with
major changes in particles' angular momentum.

\changed{Figure \ref{fig:toy_model_profile_ev} demonstrates that a strong bar can cause
a mass excess at large radii.
However, the model does not reproduce a Type-III profile as in the simulations.
In this toy model the profile looks more like a Type-II.o-OLR profile with a bump that can be
associated with the \olr of the bar \citep{Erwin2008,Gutierrez2011}.
The mass deficiency inside the \olr compared to the Type-III disk profiles in the simulations
may be due to the simplistic setup of this model.
Due to the lack of self-gravity of the particles, inward migration is likely to be
underestimated in this model.
Furthermore, we do not model star formation processes here, which are still going on in a real
disk and are strongest in the center.}

We now explore where these excess particles in the strong model come from.
Fig. \ref{fig:change_ang_mom} shows the change of specific angular momentum of each test
particle in the strong model as a function of its initial specific angular momentum.
There is a strong feature that intersects $\Delta j=0$ at the \corot.
A second strong feature corresponds to the \olr but does not intersect $\Delta j=0$.
Generally there is only moderate inward migration ($\Delta j<0$).
This can be associated with the \corot.
Since we can associate the two main features with the resonances we conclude that they are
the main drivers of the outward migration.
It is not obvious from this figure which of the resonances is dominant.

To determine which of the resonances is dominant we perform a modified resimulation of the
strong model.
We change the initial radial distribution of test particles to be a normal distribution
centred around the two resonances, \corot and \olr with a standard deviation of 0.2 kpc.
We sampled $10^5$ particles in each case.
All other parameters are identical to those given in table \ref{tab:toy_model_pars}.

\begin{figure}
    \includegraphics[width=\columnwidth]{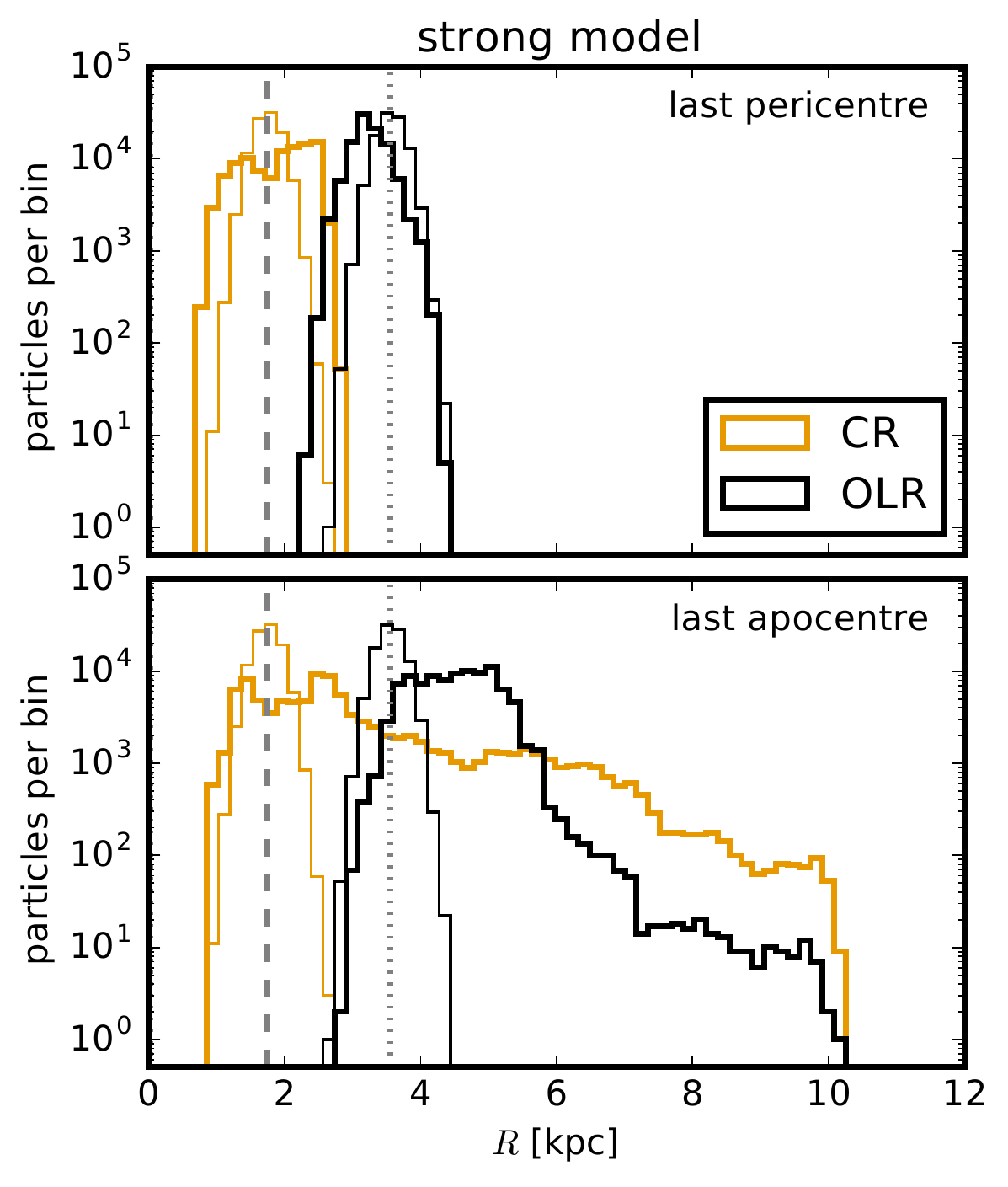}
    \caption{
        The distribution of last peri- (top panel) and apocentres (bottom) for the strong model.
        The thin histograms show the initial radial distribution of particles on a
        circular orbit centred around the \corot (orange) and the \olr (black).
        The corresponding thick histograms show the respective distribution of the
        particles' last peri- (top) and apocentre (bottom).
        For a significant fraction of the particles the apocentre is much larger than
        the pericentre indication significant radial heating.
        At intermediate radii (4-6 kpc) the heating is dominated by the \olr, at larger radii
        by the \corot.
    }
    \label{fig:last_peri_apo}
\end{figure}

Fig. \ref{fig:last_peri_apo} shows histograms of the last peri- (top panel) and the last
apocentre distance (bottom) of the test particles.
The thick orange and black histograms correspond to the particles centred around the \corot
and \olr respectively.
The thin histograms show the distribution of the corresponding initial radii.
The figure shows that the pericentre positions of the particles do not significantly differ from
the particles' initial radii.
A large fraction of the particles' apocentre distances, however, evolve to quite large values.

At intermediate radii (4-6 kpc) particles initially at the \olr outnumber the particles
initially at \corot.
This excess is however biased as the particle density at the \olr and \corot are the same
in our toy model.
In real galaxies, however, the particle density decreases with radius.
For realistic values of the disc scale length the particle density at the \olr is overestimated
by no more than a factor of 3.
Thus, correcting for this the particles in resonance with the \olr will still dominate over
those from the \corot at these radii.
At larger radii the distribution of the last apocentre distances is clearly dominated by
particles originating from the \corot.
We conclude that the excess of particles in the outer disc is caused by particles in
resonance with the \corot and the \olr.
However, the \corot is more efficient at scattering particles to very large radii.

Fig. \ref{fig:last_peri_apo} also shows that a lot of particles of the strong simulation
that are in \corot with the perturbation
have apocentre distances which are much larger than their pericentres.
This means that they experience severe radial heating indicating that the approximation
of \tsb (equation \ref{eq:heating_radial_migration}) breaks down for such a strong
perturbation.
This is also evident in the altered radial surface density profile (Fig.
\ref{fig:toy_model_profile_ev}).
\tsb showed that migrating particles in resonance with a weak perturbation exchange orbits
with another particle resulting in an unchanged global profile.
In the case of our toy models, only the weak case seems to agree with this result of \tsb.

We note that in our idealized toy model all the orbits stay bound to the region of
the perturbation.
As long as the perturbation persists, they may also return to their original orbit.
In fact the radial position of the particles show a beat pattern.
These are however not synchronous among the different particles such that the global radial
distribution of the particles will eventually reach an equilibrium distribution.
It takes about a Hubble time for such an equilibrium distribution to be reached in our toy
model.

In the simulations the scattered outer disc stars also stay bound to the bar
(top panel in Fig. \ref{fig:star_orbit}).
The situation in the simulations is more complex though, \eg the angular momentum changes even
when the stars are far away from the bar region.
As a result the star particles' orbits do not show a clear beat pattern.
A consequence is that the star's usually never return to their birth orbit.

\subsection{Toy model summary}
\label{sec:toy_model_summary}

We demonstrate here that a strong, non-axisymmetric, steady, and rotating
perturbation can cause significant radial heating at the \corot.
As a consequence the radial density profile of a disc can change significantly for such
a strong perturbation.
This is in qualitative agreement with the properties of the low spin simulation
($\lambda=0.02$) we analysed in section \ref{sec:results}.
For the weak model there is no significant change in the radial profile and only very little
heating, even at the \olr.
In this case the approximation of no radial heating at the \corot holds and it agrees well
with the results of \tsb.

Since we treat particles as test particles and ignore their self-gravity, our toy model
is not self-consistent.
Due to the way we set up the potential perturbation the additional
potential well of the perturbation at the \olr is shallower than that at the \corot.
This may limit the extent of radial heating caused by the \olr and thus it is most likely
underestimated.

Despite the possible underestimation of radial heating by the \olr, we have shown
here that particles in \corot or \olr with a strong rotating perturbation in the centre
experience significant radial heating.
It is strong enough to alter the radial surface density profile of an initially exponential
disc.
This is incompatible with the results of \tsb for weak perturbations which do not
significantly alter the global profile of their simulated discs.

In our case of a strong bar, however, we violate the approximation of a weak
perturbation \psb.
Hence, the result that particles in \corot do not experience radial heating
no longer holds.
This approximation was based on the fact that at the \corot the slope of the trajectory
in the Lindblad diagram is equal to that of the circular velocity curve (Fig. 1 in \tsb).
If the change in angular momentum due to the interaction with the perturbation is strong
enough, the particle may depart from the circular velocity curve (or increase its distance
to that).
This results in radial heating and appears to be what happens in our strong model and
in our simulations that produce Type-III disc breaks,
given that these are the simulated galaxies with strong bars (Fig. \ref{fig:bar_strength}).

Due to the lack of self-consistency and the non-axisymmetric external potential,
global angular momentum of the test particles is not necessarily conserved (see Fig.
\ref{fig:change_ang_mom}).
This may lead to an underestimation of inward migration in the toy model \changed{and might
be the reason for the mass deficiency in the central regions of this toy model}.
The model does still demonstrate the capability of such a perturbation to alter the global
mass distribution of a stellar disc.

\section{Observational signatures}
\label{sec:observations}

In this section we will give some observational signatures of the 
mechanism that we described above.
We will compare the disc dynamics of a simulated Type-III disc
(low-spin, $\lambda=0.02$) and a Type-II disc (high-spin, $\lambda=0.06$).

We have shown that the eccentricity distribution of the outer discs in the
simulations with Type-II and III disc breaks is significantly different (\cf section
\ref{sec:outer_discs}).
This leads to distinct dynamical properties of the discs that can be observed.

\begin{figure}
\includegraphics[width=\columnwidth]{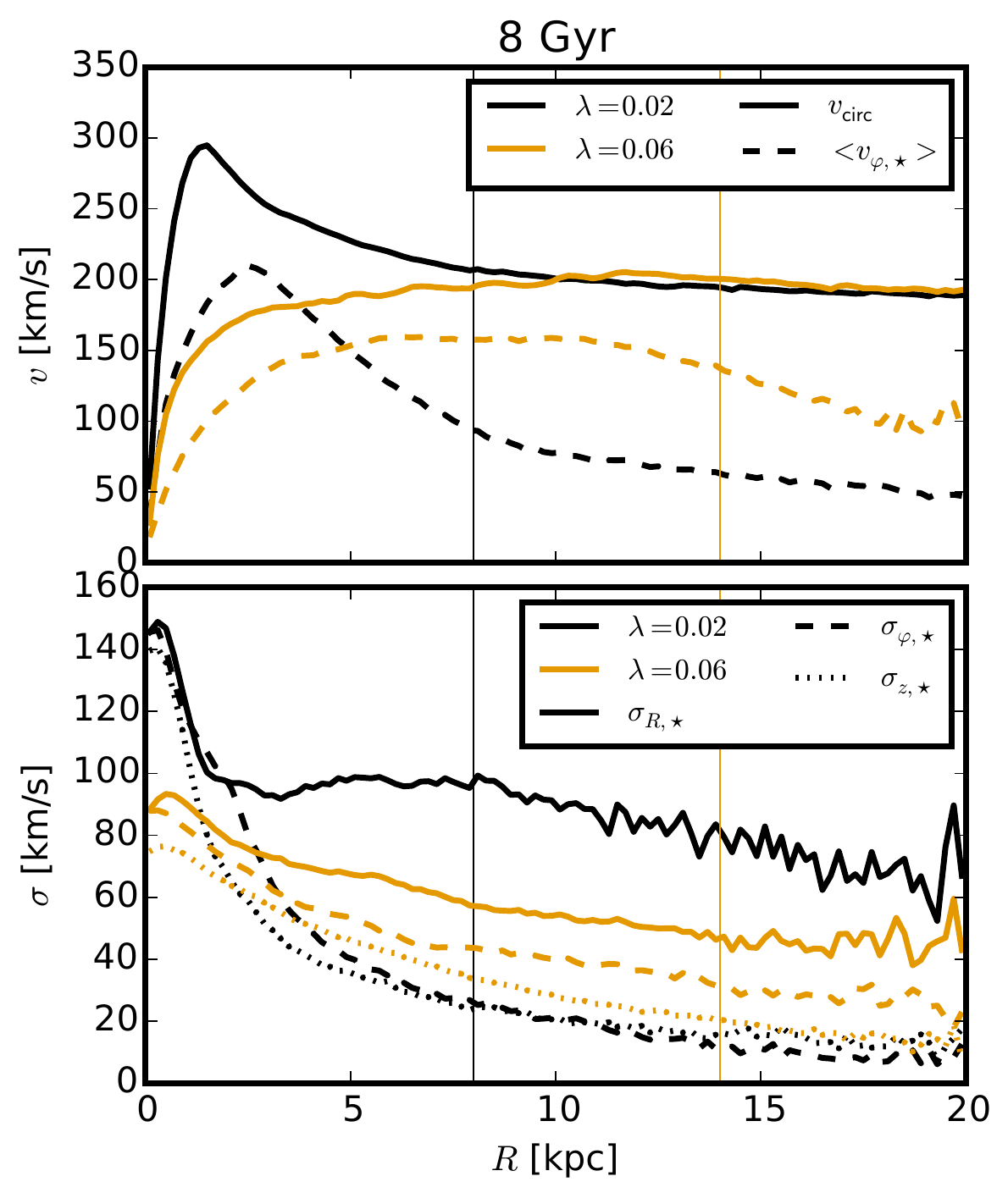}
\caption[Rotation velocity and velocity dispersion profiles]{
    Rotation velocity and velocity dispersion profiles.
    The upper panel shows the mean tangential velocity of stars (dashed lines)
    in annular bins
    and the circular velocity curve (solid lines).
    The lower panel shows the velocity dispersion in radial (solid), tangential (dashed) and
    vertical (dotted) direction.
    The vertical lines indicate the location of the breaks.
}
\label{fig:rotcurves}
\end{figure}

The upper panel of Fig. \ref{fig:rotcurves} shows the mean tangential velocity profile
$\left<v_\varphi\right>\left(R\right)$ of the
stellar discs (dashed lines) for the low-spin ($\lambda=0.02$) and high-spin
($\lambda=0.06$) galaxies.
They are compared to the expected tangential velocities for circular orbits (solid lines).
As the stellar discs host stars on non-circular orbits, we expect the mean tangential
velocity to be smaller than the circular velocity as stars at
apocentre have low velocities.
The effect is much stronger for the low-spin galaxy which exhibits
a discrepancy by a factor of 4 in the entire outer disc.
For the high-spin galaxy, the mean tangential velocity is only off by a
factor of up to 2.
This trend of decreasing tangential velocities in the outer discs with decreasing spin
is expected as the distribution of circularity parameters shifts towards more eccentric
orbits for the low-spin simulations.

The difference between the tangential velocity profile and the circular velocity curve
is difficult to observe, as the circular velocity curve cannot be observed easily.
A better diagnostic for both types of disc breaks (Type-II and III)
is the velocity dispersion profile as these can be observed.
The velocity dispersion profiles in each direction are presented in the lower panel of
Fig. \ref{fig:rotcurves}.
The prominent difference between both simulations is that the radial velocity dispersion
in the $\lambda=0.02$ (Type-III profile) disc is much higher than the tangential
and vertical component.
This is a signature of the large amount of eccentric orbits in these discs.
In the disc with a Type-II break, all velocity dispersion components are
comparable as it hosts a much smaller number of stars on eccentric orbits.

The resulting Type-III disc is dynamically peculiar.
While it has as disc shape (Fig. \ref{fig:sunrise}), it is very slowly rotating and
has an excess in radial velocity dispersion and, thus, the rotational contribution
to support the disc is unusually small.

Observing these signatures is challenging as the outskirts of Type-III discs
are very faint ($>27\,\mathrm{mag\,arcsec}^{-2}$).
Thus, they are not observable through stellar absorption-line spectroscopy, even with
integral field spectroscopy.
A possible way to measure the kinematics of the outskirts of such discs would be to
observe planetary nebulae.
Planetary nebulae have been used previously to measure velocity dispersions of galaxies
at very low surface brightness \citep[\eg][]{Douglas2000, Herrman2009} and, thus, would
provide a possible tool to observe an excess of radial velocity dispersion,
the smoking gun of stars surfing in the bar's potential.

\section{Discussion}
\label{sec:discussion}

We have presented a thorough analysis of stellar orbit evolution in one of the simulated
discs of \thfifteen{} with a Type-III break.
The goal was to identify a mechanism that produces Type-III breaks via secular
evolution.
We chose a simple simulation set-up of disc formation within non-cosmological isolated halos
to have full control over the halo spin and to facilitate the identification of secular,
\ie internal, mechanisms.

We showed that the stars that make up the outer parts of these Type-III discs were
born on circular
orbits well inside the final break radius (Fig. \ref{fig:rxy_hist} and \thfifteen).
Furthermore, the stars in the outer parts of these discs have very eccentric
orbits.
Therefore, the mechanism that produces Type-III discs is inconsistent with that
found by \citet{Roskar2008}.
They found that the stars that populate the truncated region of Type-II discs have been
churned there
from their birth radii while retaining near-circular orbits.
Instead, there has to be a mechanism that is able to drive stars to larger radii and that can
turn those stars' orbits from circular to eccentric in our low-spin simulations.

We investigated the bar as a potential cause for the formation of Type-III profiles
because the two simulations that form Type-III discs have significantly stronger
bars than all other simulations of our sample (Fig. \ref{fig:bar_strength}).
We find that the
increase of orbital radii (or more accurately their semi-major axes) is driven by an increase
in orbital energy (Fig. \ref{fig:star_orbit}) which, in turn, is driven by encounters of the
stars with the bar potential at their pericentres.
Our main argument for the bar as the dominant cause is the fact that the ensemble
of orbits of stars, that make up the outer disc, show a clear signature of the bar: the
distribution of the ratio $\Delta e/\Delta j_z$ peaks at the value of the bar's pattern speed
(Fig. \ref{fig:de_dj_hist}).
This is a clear signature that is expected from a rotating bar (equation \eqref{eq:delta_e},
see also \tsb), with a constant pattern speed (Fig. \ref{fig:pattern_speed}).

\citet{Roskar2008} found that radial migration in simulated galaxies with Type-II
breaks is dominated by churning of stars on near-circular orbits.
This preserves the stars' circularity and thus does not cause significant radial heating.
We found those results to be consistent with our simulations with Type-II profiles.
In our low-spin galaxies, which feature strong bars, the stars in \corot with the bar
may experience radial heating.
This is possible because the churning mechanism is based on a first order approximation
for moderate perturbations which states that stars in \corot do not experience radial
heating.
This first order approximation breaks down for the strong bars in our simulations.
Here the stars may be driven away from the circular orbit curve while in \corot with
the bar.
This effect of bars has been known for decades \citep{Hohl1971} but has never been linked
to the formation of Type-III stellar disc profiles.

We confirmed our interpretation of the simulations with the help of a simple
toy model which qualitatively mimics the situation in the simulations.
We showed that a strong bar-like perturbation may scatter stars to large radii in a very
similar fashion as the bar in our low-spin simulations.
An excess of stars at large radii could be reproduced in this model with a strong perturbation
but not in a model with a weaker perturbation (Fig. \ref{fig:toy_model_profile_ev})
which is in qualitative agreement with the results from our simulations.
\changed{However, the detailed shape of the final profile in the toy model does not feature
at Type-III break which is due to the lack of modelling the full complexity of galaxy evolution.}

As of now the literature only provided models that associate environmental effects
(satellites, accretion) with the formation of Type-III discs.
As this work presents a mechanism that is capable of producing these discs in simulations
of isolated galaxies, we show that galaxies can form Type-III discs purely via secular
evolution.
This does not preclude that external forces \citep{Younger2007, Kazantzidis2009, Roediger2012,
Borlaff2014} or other internal processes may also play a role.

We find that the initial halo spin matters as it locally sets the rate at
which stars form.
This in turn determines the size of the stellar disc and its central surface density.
Hence, it also sets
the strength and longevity of any bar at the centre.
This effect cannot be explored in simulations with prepared discs \citep[\eg][]{Foyle2008}.
For high spin parameters the gaseous disc is very extended but the gas surface density is
comparatively low.
This leads to more extended star formation at a rather low rate.
In low spin halos the opposite happens.
Due to the large amount of low angular momentum gas a large portion of this gas is able to
settle at low radii forming a concentrated massive gas disc with a high star formation rate.
In this case a massive and concentrated disc forms which is highly susceptible to disc
instabilities and thus very prone to bar formation \citep{Toomre1964, Hohl1971}.
Since the stellar mass density in the centre is very large, the bar is also very massive.
As we have shown in this work, this massive bar is driving the formation of 
Type-III discs.

We previously reported a correlation between disc scale length and the type of truncation
(middle panel of Fig. 2 in \thfifteen), \ie the inner scale lengths increase from compact
discs for type-III profiles through intermediate scale lengths for type-I profiles and
extended discs for type-II profiles.
While \citet{Maltby2012} and \citet{Head2015} do not find such a correlation in their
galaxy samples, \citet[section 5.2]{Gutierrez2011} report a trend that qualitatively
agrees with our simulations \phfifteen.
\citet{Gutierrez2011} cannot rule out a common distribution for the inner scale lengths
of type-I and II discs but find that Type-III discs have significantly
shorter inner scale lengths compared to pure exponentials.
Despite this qualitative agreement between a subset of observational data and our simulated
galaxies, it is not the goal of this project to reproduce previous observational results but to
take a first step to a possible explanation of the formation of Type-III discs
which relies purely on secular stellar dynamics.

We stress that we do not claim the proposed mechanism to be exclusively responsible for
the existence of Type-III discs.
Instead we show that the proposed mechanism is potentially capable of producing
Type-III discs and characterise its dynamical signature which can be used to
observationally verify or falsify our model.

The eccentric orbits of the stars in the outskirts of Type-III
give rise to very unusual properties of a stellar disc:
While being flat the outer disc rotates very slowly and its velocity dispersion is
dominated by the radial component.

\citet{Minchev2012} find similar kinematic signatures in Type-III
discs that form through external gas accretion.
This degeneracy does not undermine the potential capability of falsifying our model if
no Type-III discs with bars can be found that have a strong excess in radial
velocity dispersion in their outskirts.

We showed that stars in the outskirts of simulated Type-III discs
are on very eccentric orbits.
They were formed on circular orbits at much smaller galactocentric radii.
This outward migration was caused by a strong bar that formed 3-4 Gyr into the
simulation and trapped the migrating particles in its rotating potential.
This caused a change of the trapped stars' radial distribution which is responsible
for the formation of the anti-truncated part of Type-III discs.
This conclusion is based on a dominant signature of the bar on stellar orbits.
We describe signatures of this mechanism that allow to test our predictions in
future observations.

\section{Conclusion}
\label{sec:Conclusion}

Analyzing controlled disc-galaxy simulations we showed that strong bars can 
form anti-truncated breaks in the radial surface density profile of stellar discs (Type-III
discs).
Strong bars can boost the semi-major axis of stellar
orbits to very large galactocentric distances, while also changing
the distribution of orbit circularity of the affected stars.
The initial distribution of stars that later comprise the outskirts of a simulated Type-III
stellar disc was initially dominated by near-circular orbits at smaller radii.
In the final distribution, near-circular orbits are almost completely absent.
It is dominated by very eccentric orbits.
This bar-induced transformation of orbits has the capability of changing the global radial
mass distribution of the stellar disc ultimately leading to a Type-III break in
the radial surface-density profile.

This process is not efficient for discs with weak bars.
In this case the dominant radial migration mechanism is spiral-induced churning and the
discs form Type-II breaks in their radial profile.
The reason is that the first order approximation of only very little
or no radial heating holds only for weak but not for strong bars.

\section*{Acknowledgements}
The authors gratefully acknowledge the very constructive help of various people, in particular
Rok Ro\v skar, Victor Debattista, Ivan Minchev, Deidre Hunter and Francoise Combes.
We thank the anonymous referee for their comments which helped us to improve upon the
interpretation of our results.
This paper could not have happened without the $N$-body hydrodynamics code {\sc ChaNGa} whose
development has been shepherded by Thomas Quinn and whose hydrodynamics come from the work of
James Wadsley. We thank them for their contributions and letting us use their code.
JH, GSS and HWR acknowledge funding from the European Research Council under the
European Union's Seventh Framework Programme (FP 7) ERC Advanced Grant Agreement n. [321035].
GSS and AAD acknowledge support through the  Sonderforschungsbereich SFB 881
``The Milky Way System'' (subproject A1) of the German Research Foundation (DFG).
MM acknowledges support from the Alexander von Humboldt Foundation.

\input{herpich17b.bbl}

\end{document}